\documentclass[aps,prl,groupedaddress,showpacs]{revtex4}
\usepackage{graphicx}
\usepackage{float}
\usepackage{dcolumn}
\usepackage{amsmath}
\usepackage{amssymb}
\input{epsf}

\def\be{\begin{equation}}
\def\ee{\end{equation}}
\def\bea{\begin{eqnarray}}
\def\eea{\end{eqnarray}}

\def\half{1 \over 2}

\newcommand{\ba}{\begin{eqnarray}}
\newcommand{\ea}{\end{eqnarray}}
\newcommand{\x}{\mbox{$\vec{x}$}}
\def\ket#1{\vert #1 \rangle}
\def\bra#1{\langle #1 \vert}
\def\bracket#1#2{\langle #1 \vert #2 \rangle}
\def\vev#1{\left< #1 \right>}

\newcommand{\Adag}{{A^{\dag}_\sigma}}
\newcommand{\A}{{A_\sigma}}

\newcommand{\Schrodinger}{Schr\"odinger\ }

\begin{document}

\title{
\noindent\hfill\hbox to 1.5in{\rm  } \vskip 1pt \noindent\hfill\hbox
to 1.5in{\rm SLAC-PUB-13759 \hfill  } \vskip 1pt
\vskip 10pt
Dynamic quantum clustering: a method for visual exploration of structures in data \footnote{This work was
supported by the U.~S.~DOE, Contract No.~DE-AC02-76SF00515.}}
\author{Marvin Weinstein$^1$ and David Horn$^2$\\
\normalsize{$^{1}$Stanford Linear Accelerator Center, Stanford, CA, USA}\\
\normalsize{$^{2}$School of Physics and Astronomy, Tel Aviv University, Tel Aviv 69978, Israel}}

\begin{abstract}
A given set of data-points in some feature space may be associated with a \Schrodinger
equation whose potential is determined by the data. This is known to lead to good
clustering solutions. Here we extend this approach into a full-fledged dynamical scheme
using a time-dependent \Schrodinger equation. Moreover, we approximate this Hamiltonian
formalism by a truncated calculation within a set of Gaussian wave functions (coherent
states) centered around the original points. This allows for analytic evaluation of the
time evolution of all such states, opening up the possibility of exploration of
relationships among data-points through observation of varying dynamical-distances among
points and convergence of points into clusters. This formalism may be further
supplemented by preprocessing, such as dimensional reduction through singular value
decomposition or feature filtering.
\end{abstract}
\maketitle
\section{Introduction}

Clustering of data is, in general, an ill-defined problem.
Nonetheless it is a very important one in many scientific and technological
fields of study. Given a set of data-points one looks for possible structures
by sorting out which points are close to each other and, therefore, in some sense
belong together. This is a preliminary stage taken before investigating what
properties are common to these subsets of the data.

It has recently become quite popular to investigate such questions
in a dynamic framework, thus allowing for more fluid associations of
data-points, rather than rigid clusters. Diffusion geometry
\cite{coifman, lafon, nadler} is such a method, based on a discrete
analog of the heat equation
\be
- {\partial \Phi \over \partial t} = H \Phi
\ee
where $H$ is some operator with positive eigenvalues, guaranteeing that
the temporal evolution of $\Phi(\x,t)$ is that of diffusion. Thus, starting
out with $\Phi(\x,0)$, e.g. a Gaussian concentrated around some data point
one would expect  $\Phi(\x,t)$ to spread over all space that is occupied by
the data points.

In this paper we advocate the use of a Schr\"odinger Hamiltonian $H$
that is intimately connected to the data-structure, as defined by
the quantum clustering method \cite{qc1} and summarized below. We
extend it into a time-dependent Schr\"odinger equation:
\be
{i} {\partial \psi(\vec{x},t) \over
\partial t} = H \psi(\vec{x},t)
\ee
The ensuing Dynamic Quantum Clustering (DQC) formalism allows
us, by varying a few parameters, to study in detail the temporal
evolution of wave-functions representing the original data-points.
In turn, this dynamical behavior allows us to explore the structure
of the quantum potential function defined by the quantum clustering
method.

DQC begins by associating each data-point with a state in
Hilbert space. The temporal development of the centroids of these
states may be viewed in the original data-space as moving images of
the original points. Their distances to one-another change with
time, thus representing associations they form with each other.
Convergence of many points onto a common center at some instant of
time is an obvious manifestation of clustering. Many transitional
relationships may occur, revealing substructures in clusters or even
more complex associations. For this reason we propose this approach as a
general method for visually and interactively searching for and
exploring structures in sets of data.

\section*{Quantum Clustering}

The quantum clustering approach begins, as does the well-known
Parzen-window estimator\cite{duda}, by associating to each of $n$
data points $\x_i$ in a Euclidean space of $d$ dimensions a Gaussian
wave-function $ \psi_i(\x) =  e^{-{(\x -\x_i)^2 \over 2\sigma^2}} $
and then constructing the sum of all these Gaussians, \be \psi(\x) =
\sum_i e^{-{(\x -\x_i)^2 \over 2\sigma^2}}. \label{wavefn} \ee
Conventional scale-space clustering \cite{roberts} views this
function as a probability distribution (up to an overall factor)
that could have generated the observed points, and regards therefore
its maxima as determining locations of cluster centers. Often these
maxima are not very prominent and, in order to uncover more of them,
one has to reduce $\sigma$ down to low values where the number and
location of the maxima depend sensitively upon the choice of
$\sigma$.

Quantum clustering took a different approach, requiring $\psi$ to be
the ground-state of the Hamiltonian \be \label{sch} H\psi \equiv (
-{\sigma^2 \over 2}\nabla ^2 + V({\bf x}))\psi = E_0 \psi. \ee By
positing this requirement \cite{qc1}, the potential function $V({\bf
x})$ has become inextricably bound to the system of data-points,
since $V(\x)$ is determined, up to a constant, by a simple algebraic
inversion of Eq.\ref{sch}. \endnote{For a single data-point $V(\x)$
is just a quadratic harmonic potential centered on the original data
point.}  Moreover, one may expect $V$ to have minima in regions
where $\psi$ has maxima and furthermore, that these minima will be
more pronounced than the corresponding maxima found in the Parzen
estimator.  In fact, it frequently turns out that a concentration of
data-points will lead to a local minimum in $V$, even if $\psi$ does
not display a local maximum.  Thus, by replacing the problem of
finding maxima of the Parzen estimator by the problem of locating
the minima of the associated potential, $V(\x)$, we simplify the
process of identifying clusters.  The effectiveness of this approach
has been demonstrated in the work by Horn and Gottlieb\cite{qc1,
compact}.  It should be noted that the enhancement of features
obtained by applying Eq.\ref{sch} comes from the interplay of two
effects: attraction of the wave-function to the minima of $V$ and
spreading of the wave-function due to the second derivative (kinetic
term).  This may be viewed as an alternative model to the
conventional probabilistic approach, incorporating attraction to
cluster-centers and creation of noise, both inferred from - or
realized by - the given experimental data.

DQC drops the probabilistic interpretation of $\psi$ and replaces it
by that of a probability-amplitude, as customary in  Quantum
Mechanics. DQC is set up to associate data-points with cluster
centers in a natural fashion.  Whereas in QC this association was
done by finding their loci on the slopes of $V$, here we follow the
quantum-mechanical temporal evolvement of states associated with
these points.  Specifically, we will view each data-point as the
expectation value of the position operator in a Gaussian
wave-function $ \psi_i(\x) = e^{-{(\x -\x_i)^2 \over 2\sigma^2}} $;
the temporal development of this state traces the association of the
data-point it represents with the minima of $V(\x)$ and thus, with
the other data-points.

\section*{Dynamic Quantum Clustering (DQC)}

As we already noted, the conversion of the static QC method to a
full dynamical one, begins by focusing attention on the Gaussian
wave-function, $ \psi_i(\x) =C\, e^{-{(\x -\x_i)^2 \over 2\sigma^2}}
$, associated with the $i^{\rm th}$ data point, where $C$ is the
appropriate normalization factor. Thus, by construction, the
expectation value of the operator $\vec{x}$ in this state is simply
the coordinates of the original data point; i.e., \
\be
\vec{x}_i =
\bra{\psi}\,\vec{x} \, \ket{\psi} = \int d\vec{x}\,
 \psi^\ast_i(\vec{x})\,\vec{x}\,\psi_i(\vec{x}) .
\ee
The dynamical part of the DQC algorithm is that, having constructed
the potential function $V(\x)$, we study the time evolution of each
state $\psi_i(\x)$ as determined by the time dependent \Schrodinger equation;
i.e.,
\be
\label{Schdeqn}
i {\partial \psi_i(\vec{x},t) \over
\partial t} = H \psi_i(\vec{x},t) = \left(-{\nabla^2 \over 2m}
+ V(\vec{x})\right)\psi_i(\vec{x},t) ,
\ee
where $m$ is an arbitrarily chosen mass for a particle moving in $d$-dimensions
If we set $m={1 / \sigma^2}$ then, by construction, $\psi(\vec{x})$ of Eq. 3
is the lowest energy eigenstate of the Hamiltonian.  If $m$ is chosen to
have a different value, then not only does each individual state $\psi_i(\x)$ evolve
in time, but so does the sum of the states, $\psi(\x)$.

The important feature of quantum dynamics, which makes the evolution
so useful in the clustering problem, is that according to Ehrenfest's
theorem, the time-dependent expectation value
\be
 \bra{\psi(t)}\,\x\, \ket{\psi(t)} = \int\,d\x \, \psi^\ast_i(\vec{x},t)\,
\x\,\psi_i(\vec{x},t) ,
\ee
satisfies the equation,
\bea
   { d^2  \langle\,\vec{x}(t) \rangle \over dt^2} &=&
   - {1 \over m}  \int d\vec{x}
 \psi^\ast_i(\vec{x},t)\,\vec{\nabla}V(\vec{x}) \,\psi_i(\vec{x},t) \\
 &=& \bra{\psi(t)}\,\vec{\nabla}V(\vec{x})\,\ket{\psi(t)} .
\eea
If $\psi_i(\x)$ is a narrow Gaussian, this is equivalent
to saying that the center of each wave-function rolls towards
the nearest minimum of the potential according to the classical
Newton's law of motion.  This means we can explore the relation of
this data point to the minima of $V(\vec{x})$ by following the
time-dependent trajectory $\langle\,\vec{x}_i(t) \,\rangle =
\bra{\psi_i(t)}\,\vec{x}\,\ket{\psi_i(t)}$.
Clearly, given Ehrenfest's theorem, we expect to see any points
located in, or near, the same local minimum of $V(\vec{x})$ to
oscillate about that minimum, coming together and moving apart. In
our numerical solutions we generate animations which display
this dynamics for a finite time.
This allows us to visually trace the clustering of points associated with
each one of the potential minima.

In their quantum clustering paper Horn and Gottlieb successfully
used classical gradient descent to cluster data by moving points (on classical
trajectories) to
the nearest local minimum of $V(\x)$.  The idea being that
points which end up at the same minimum are in the same cluster.
At first glance it would seem that DQC replaces the conceptually
simple problem of implementing gradient descent with the more
difficult one of solving complicated partial differential equations.
We will show the difficulty is only apparent.  In fact, the
solution of the \Schrodinger  equation can be simplified considerably,
and will also allow further insights than the gradient descent method.
The DQC algorithm translates the problem of solving
the \Schrodinger equation into a matrix form which captures most of
the details of the analytic problem, but which involves $ N \times
N$-matrices whose dimension, $N$, is less than or equal to the
number of data points.  This reduction is independent of the
data-dimension of the original problem. From a computational point
of view there are many advantages to this approach.  First, the
formulas for constructing the mapping of the original problem to a
matrix problem are all analytic and easy to evaluate, thus computing
the relevant reduction is fast.  Second, the evolution process only
involves matrix multiplications, so many data points can be
evolved simultaneously and, on a multi-core processor, in parallel.
Third the time involved in producing the animations showing how the
points move in data space scales linearly with the number of
dimensions to be displayed.  Finally, by introducing an $m$ that is
different from $1/ \sigma^2$ we allow ourselves the freedom of
employing low $\sigma$, which introduces large numbers of minima
into $V$, yet also having a low value for $m$ which guarantees efficient
tunneling, thus connecting points that may be located in nearby, nearly
degenerate potential minima. By using this more general Hamiltonian, we
reduce the sensitivity of the calculation to the specific choice of
$\sigma$.

One final point worth making before describing the method of calculation
is that the use of Gaussian wave-functions to represent data points allows
us to develop a number of flexible strategies for handling very
large data sets.  This issue will be addressed below.

\section*{The Calculation Method}

Before discussing how this method works in practice we will give a
brief outline of the details of the general procedure.  We begin by
assuming that there are $n$-data points that we wish to cluster.  To
these data points we associate $n$-states, $\ket{\psi_i}$.  These
states are $n$ Gaussian wave-functions such that the $i^{\rm th}$
Gaussian is centered on the coordinates of the $i^{\rm th}$ data
point. These states form a basis for the vector space within
which we calculate the evolution of our model.

Let us denote by $N$, the $n \times n$ matrix formed from the scalar products
\be
    N_{i,j} = \bracket{\psi_i}{\psi_j},
\ee
and by ${\cal H}$, the $n \times n$-matrix
\be
   {\cal H}_{i,j} = \bra{\psi_i} H \ket{\psi_j},
\ee
and by
$\vec{X}_{i,j}$ the matrix of expectation values
\be
    \vec{X}_{i,j} = \bra{\psi_i} \vec{x} \ket{\psi_j}.
\ee

The calculation process can be described in five steps.
First, begin by finding
the eigenvectors of the symmetric matrix $N$ which correspond to
states having eigenvalues larger than some pre-assigned value; e.g., $10^{-5}$.
These vectors are linear combinations of the original Gaussians which
form an orthonormal set.  Second, compute ${\cal H}$ in this
orthonormal basis, ${\cal H}^{tr}$. Do the same for $\vec{X}_{i,j}$.
Fourth, find the eigenvectors and eigenvalues
of ${\cal H}^{tr}$, construct
$\ket{\psi_i(t)} = e^{-i t\,{\cal H}^{tr}}\, \ket{\psi}$,
that is the solution to the reduced time dependent \Schrodinger problem
\be
    i {\partial \over \partial t}\,\ket{\psi_i(t)} = {\cal H}^{tr}\,\ket{\psi_i(t)}
\ee
such that $\ket{\psi_i(t=0)} = \ket{\psi_i}$.
Finally, construct the desired trajectories
\be
\vev{\vec{x}_i(t)} = \bra{\psi_i} e^{ i t {\cal H}^{tr}}\, \vec{X} \,
e^{-i t {\cal H}^{tr}} \ket{\psi_i}
\ee
by evaluating this expression for a range of $t$ and
use them to create an animation.  Stop the animation when clustering
of points is obvious.

It is clear that restricting attention to the truncated Hamiltonian
perforce loses some features of the original problem, however its
advantage is that we can derive analytic expressions for all
operators involved (see Appendices A and B). As a result, the
numerical computations can be done very quickly.  Experience
has shown that as far as clustering is concerned this approximation
causes no difficulties.

\section*{Example: Ripley's Crab data}

To test our method we apply it to a five-dimensional dataset with
two-hundred entries, used in Ripley's text book \cite{ripley}.  This
dataset records five measurements made on male and female crabs that
belong to two different species.  This dataset has been used in the
original paper on quantum clustering by Horn and Gottlieb\cite{qc1}.
It is being used here to allow readers to compare the two techniques.
Applications to other data sets will be discussed below.
Our main motivation is to provide a simple example which exhibits
the details of the DQC method.  In particular, we wish to show that
the simplest computational scheme for implementing the general
program captures the essential features of the problem and does as
well as one can reasonably expect to do.

\begin{figure}[h!]
 \hbox to \hsize{\hss
  \includegraphics[width=3.0in]{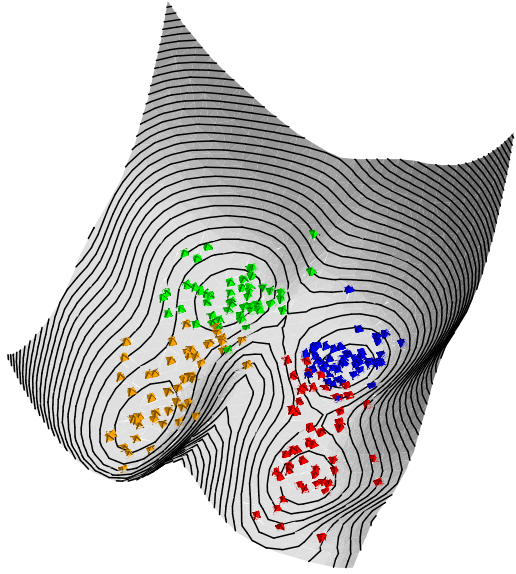}
  \hss}
  \caption{This is a plot of the quantum potential function for
  the two-dimensional problem of the Ripley's Crab Data where
  the coordinates of the data points are chosen to be given by
  the second and third principal components.  The four known
  classes of data points are shown in different colors and
  are placed upon the potential surface at their original locations.
  }
\label{twodpot}
\end{figure}

The data is stored in a matrix $M$ which has $200$ rows and $5$
columns.  Following an often-used dimensional reduction method, we preprocess
our data
with a singular-value decomposition
\be
    M = U\,S\,V^{\dag},
\ee
where $U$ is a unitary $200 \times 200$ matrix and $S$ is the $200
\times 5$ matrix of singular values, the latter occurring on the
diagonal of its upper $5 \times 5$ entries.  The sub-matrix of $U$
consisting of the first five columns, the so-called five Principal
Components (PCs), can be thought of as assigning to each sample a
unique point in a five-dimensional vector space. We may study the
problem in the full five-dimensional space, or within any subspace
by selecting appropriate principal components. In \cite{qc1} QC was
applied to this problem in a 2-dimensional subspace, consisting of
PC2 and PC3.  In what follows we will discuss the application of
DQC to the 3-dimensional data composed of the first three PCs (although
there would be no change in the results if we used all five dimensions).
In order to give the reader some feeling for how the quantum
potential associated with the data looks in a simple case, we have
included Fig.\ref{twodpot} where we exhibit the original data
points (different colors of points correspond to known classes),
placed upon the associated two-dimensional quantum potential.
As is obvious from the plot the minima of the potential function do
a very good job of capturing the different clusters.  Moreover,
letting data points roll downhill to the nearest minimum
will produce a reasonable separation of the clusters.

Clearly, when we restrict attention to the first three
PCs, the rows of the matrix obtained by restricting $U$
to its first three columns are not guaranteed to be normalized
to unity. Hence we employ the conventional approach of projecting
all points onto the unit sphere \cite{lsa}.
\begin{figure}[h!]
 \hbox to \hsize{\hss
  \includegraphics[width=2.0in]{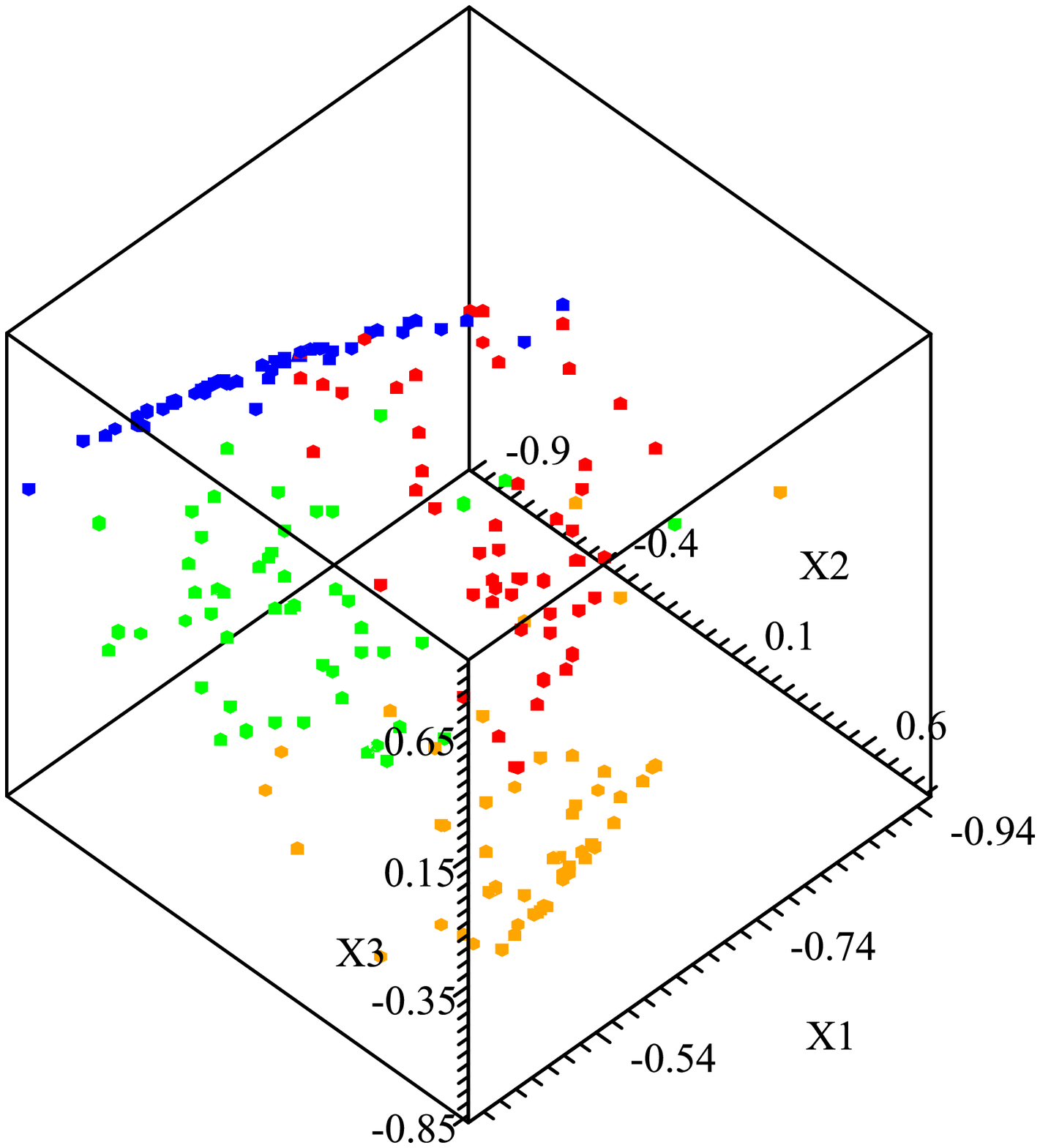}
  \hss\qquad\hss
  \includegraphics[width=2.0in]{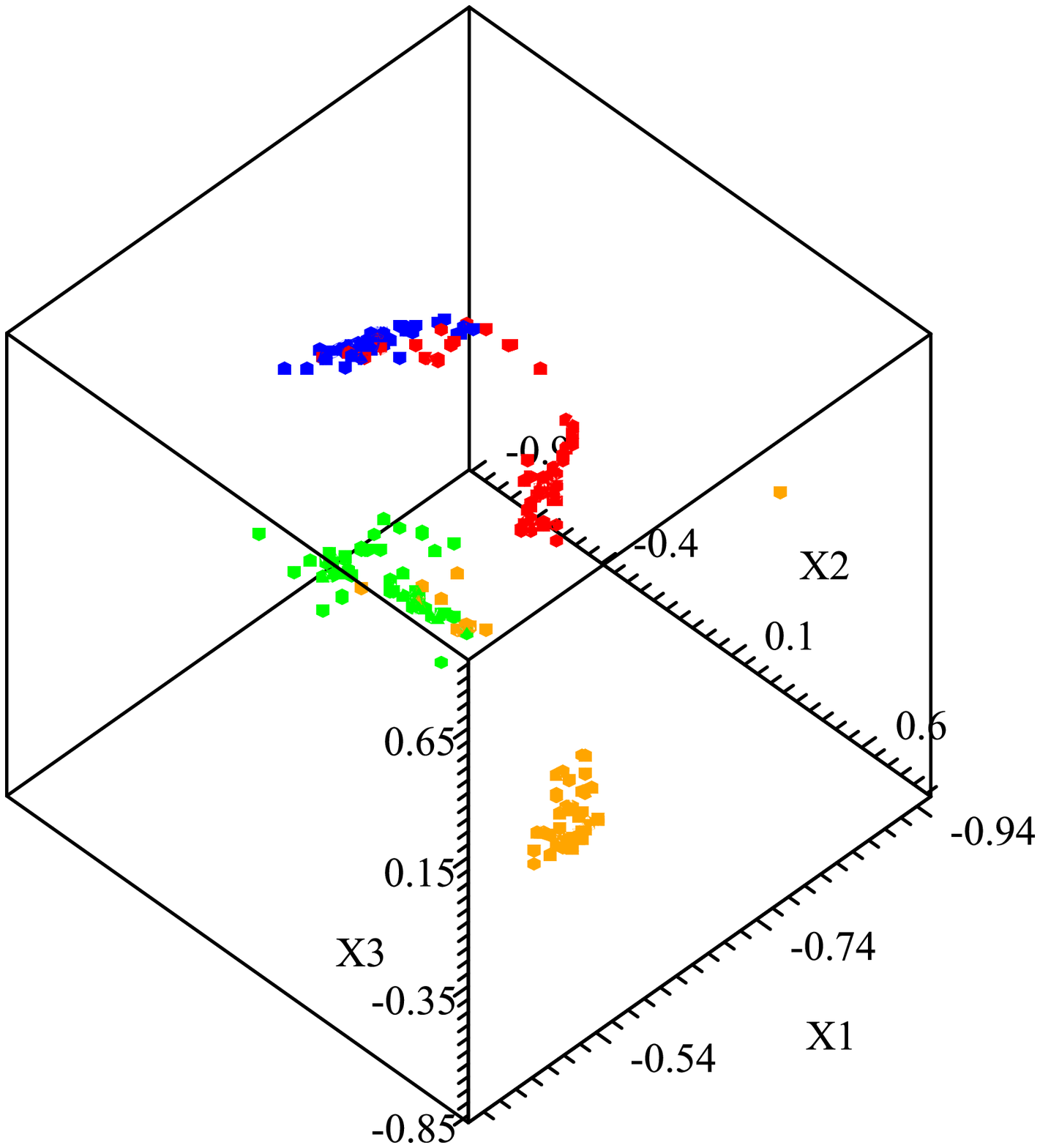}
   \hss\qquad\hss
   \includegraphics[width=2.0in]{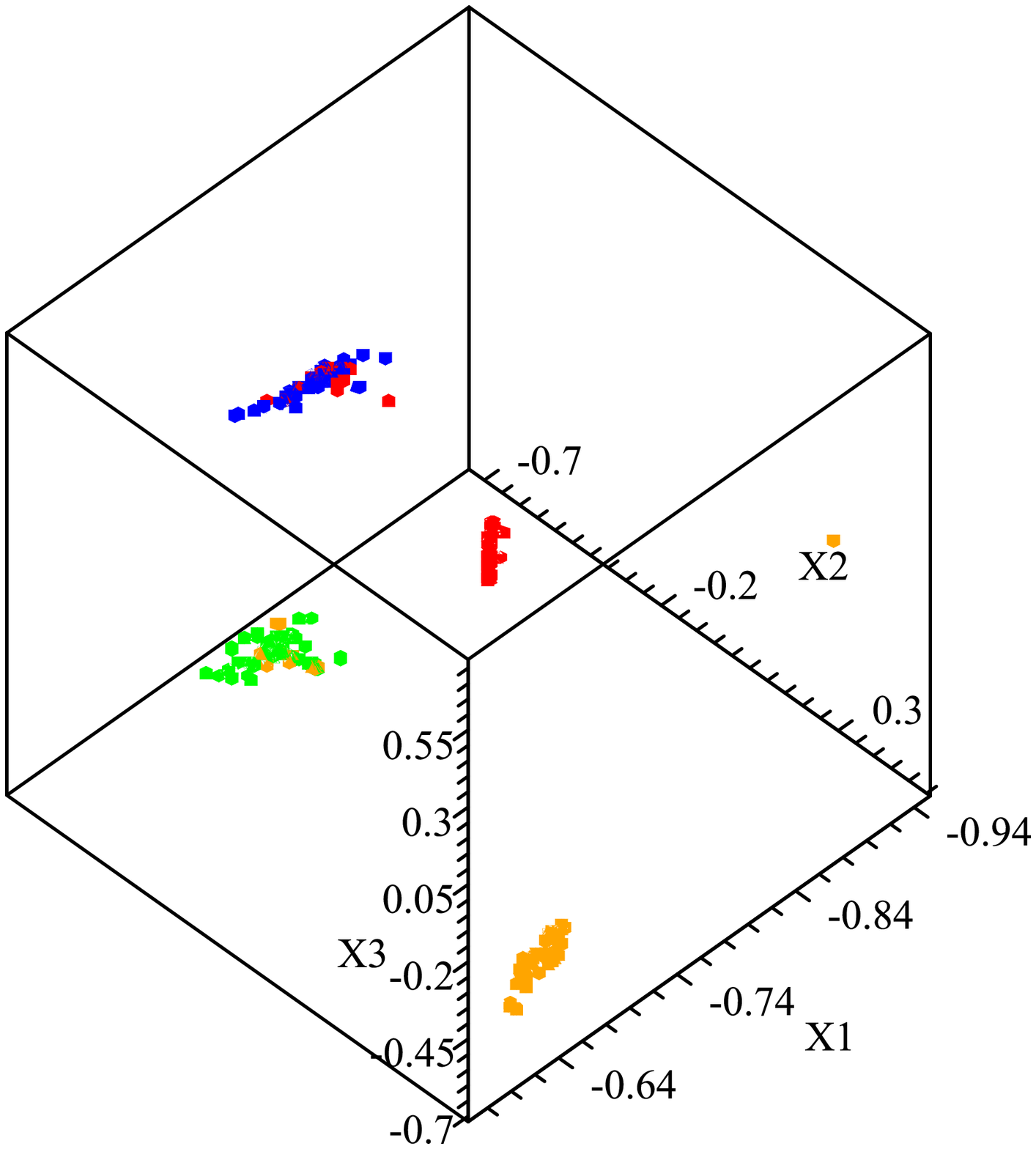}
    \hss}
  \caption{The left hand plot shows three-dimensional distribution of the
  original data points before quantum evolution.  The middle plot shows
  the same distribution after quantum evolution. The right hand plot shows the results
of an additional iteration of DQC. The values of parameters
 used to construct the Hamiltonian and evolution operator are:
  $\sigma = 0.07$ and $m = 0.2$. Colors indicate the
  expert classification of data into four classes, unknown to the clustering algorithm.
  Note, small modifications of the parameters lead to the same results.}
\label{new1}
\end{figure}

In what follows we study the temporal behavior of the curves
$\vev{\vec{x}_i(t)}$, for all $i$. Henceforth we will refer to this
as the ``motion of points''. Figure~\ref{new1} shows the
distribution of the original data points plotted on the unit sphere
in three dimensions.  This is the configuration before we begin the
dynamic quantum evolution. To visually display the quality of the
separation we have colored the data according to its known four
classes, however this information is not incorporated into our
unsupervised method.  To begin with, we see that the two species of
crabs ((red,blue) and (orange,green)) are fairly well separated;
however, separating the sexes in each species is problematic.  The
middle plot in Figure~\ref{new1} shows the distribution of the
points after a single stage of quantum evolution, stopped at a time
when points first cross one another and some convergence into
clusters has occurred.  It is immediately
obvious that the quantum evolution has enhanced the clustering and
made it trivial to separate clusters by eye. Once separation is
accomplished, extracting the clusters can be performed by eye from
the plots or by any conventional technique, e.g. k-means.

\begin{figure}[h!]
   \hbox to \hsize{\hss
   \includegraphics[width=2.0in]{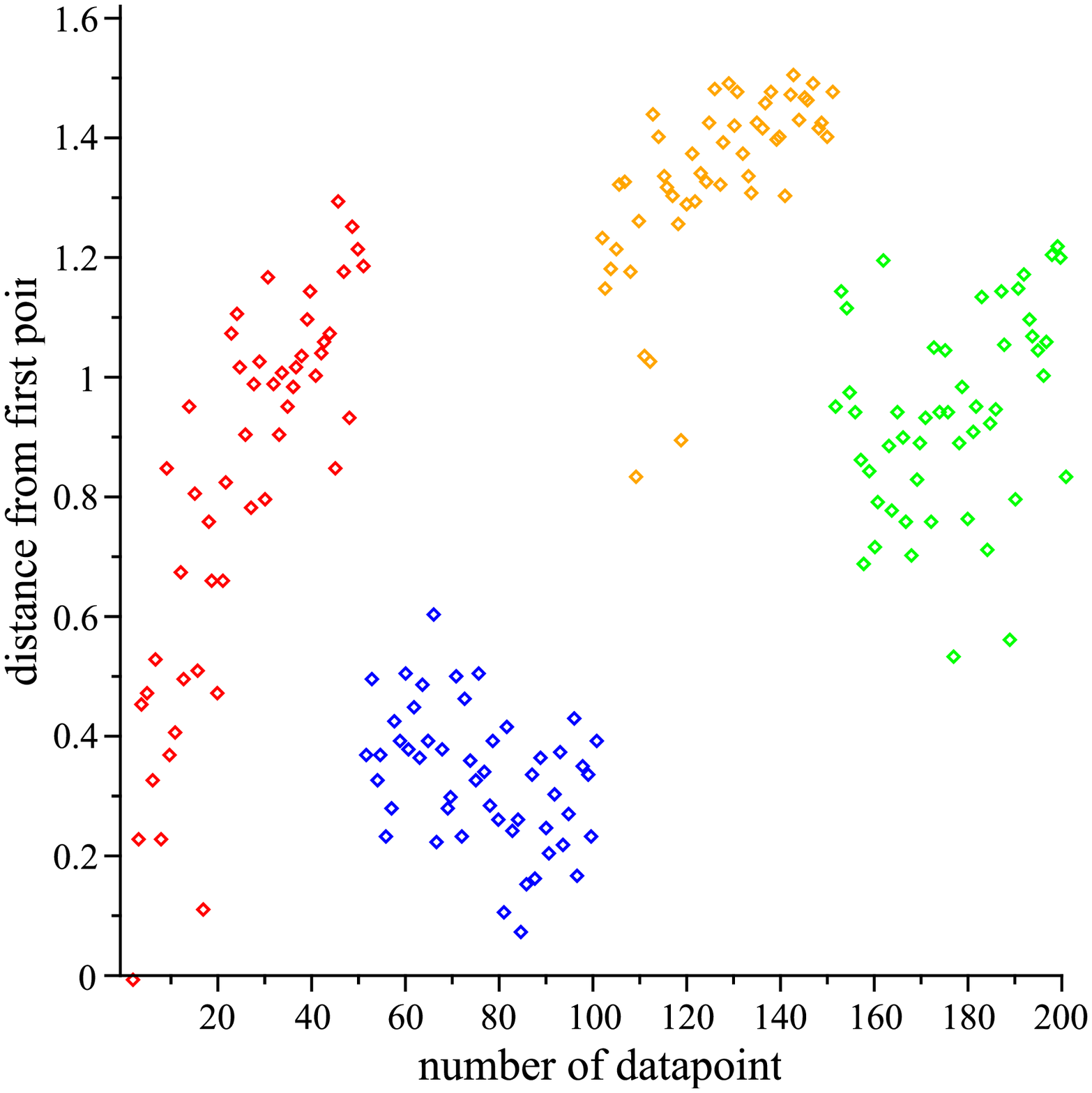}
  \hss\qquad\hss
   \includegraphics[width=2.0in]{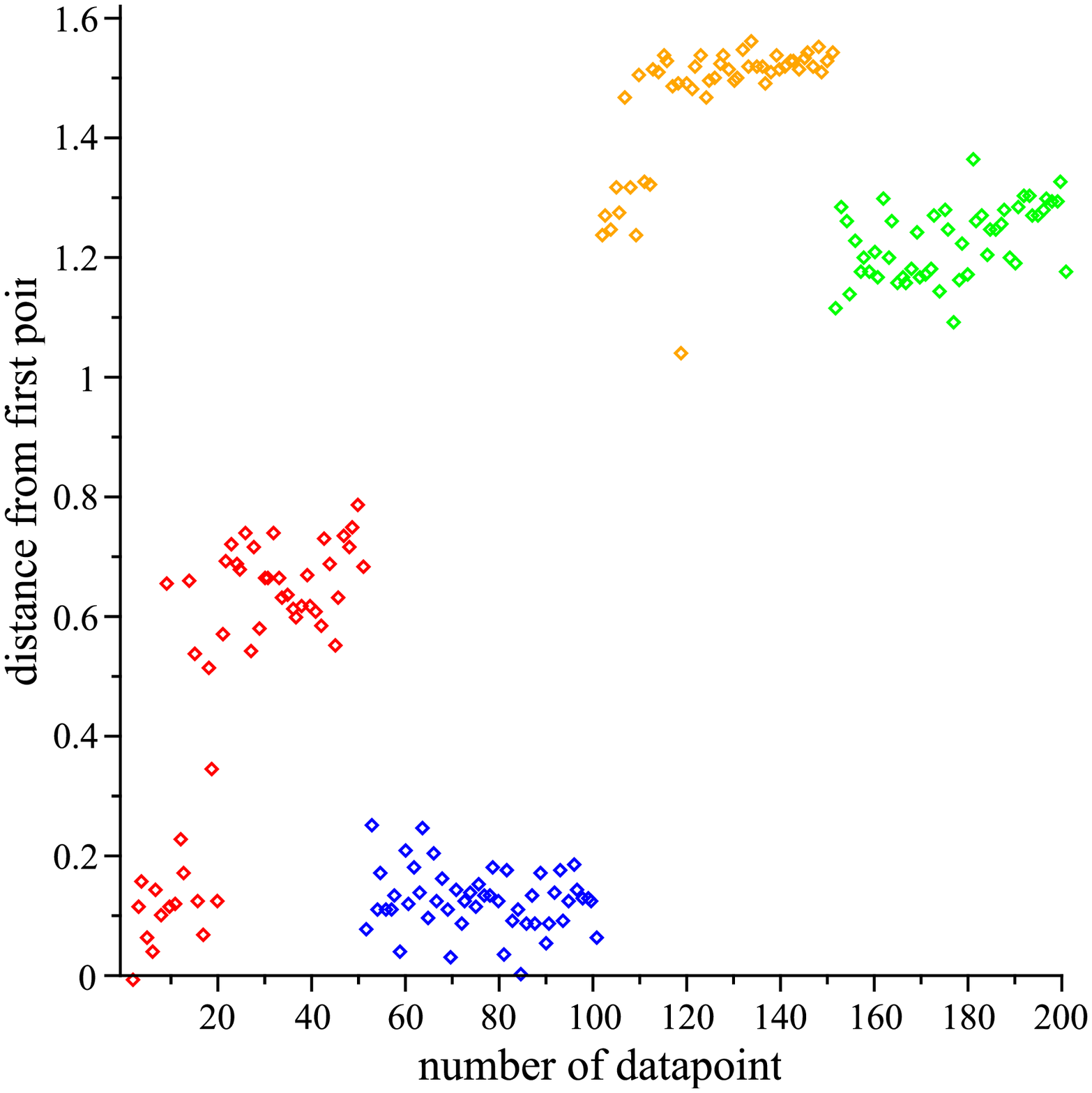}
   \hss\qquad\hss
   \includegraphics[width=2.0in]{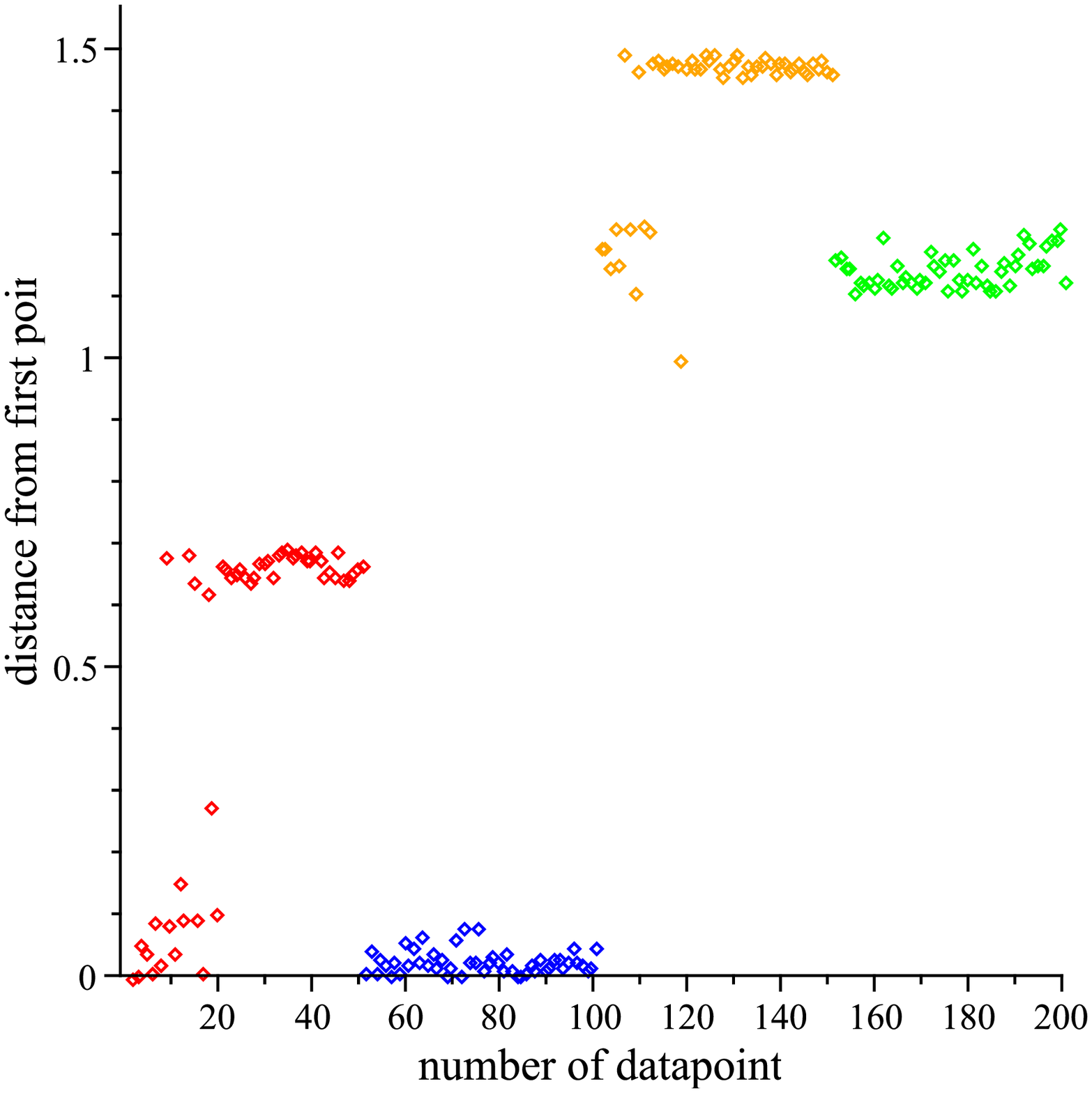}
    \hss}
  \caption{A plot of Euclidean distance of each point $i$ from the first
  data point.  Again, the left hand plot shows the distances for the initial
  distribution of points.  The middle plot shows the same distances after
  quantum evolution. The right-hand plot shows results after another iteration of DQC.
  The numbering of the data-points is ordered according to the expert classification of these points into
  four classes containing 50 instances each.}
\label{new2}
\end{figure}

An alternative way of displaying convergence is shown in
Figure~\ref{new2}, where we plot the Euclidean distance from the
first point in the dataset to each of the other points.  The
clusters lie in bands which have approximately the same distance
from the first point.
It is difficult to get very tight clusters since the points, while
moving toward cluster centers, oscillate around them, and arrive at
the minima at slightly different times. Given this intuition, it is
clear that one way to tighten up the pattern is to stop DQC
evolution at a point where the clusters become distinct, and then
restart it with the new configuration, but with the points redefined
at rest.  We refer to this as iterating the DQC evolution. The
right-hand plots in Figure~\ref{new1} and Figure~\ref{new2} show
what happens when we do this.  The second stage of evolution clearly
tightens up the clusters significantly, as was expected.

By the end of the second iteration, there can be no question that it
is a simple matter to extract the clusters. As is quite evident,
clustering does not agree completely with the expert classification,
i.e. points with different colors may be grouped together. This is,
however, the best one can do by color-blind treatment of the
information provided in the data-matrix. As we already noted the
full 5-dimensional study of the crab data-set can proceed in the
same manner, although it does not lead to new insights.

\section*{Dynamic Distances}

The fact that data-points of different classes happen to lie close
to each other in the data-matrix can be due to various factors:
errors in data measurements, errors in the expert assignment to
classes, true proximity of data-points in spite of differences of
origin (extreme example would be similarities of phenotypes in spite
of differences in genotypes) or - the simplest possibility - the
absence of some discriminative features in the feature-space that
spans the data measurements. But there is another important
conceptual message to be learned here - clustering and/or
classification may not capture all the interesting lessons that may
be derived from the data. A similar message is included in the
Diffusion Geometry approach \cite{coifman, lafon} that advocates
measuring diffusion-distances among points rather than Euclidean
ones. Diffusion distances are influenced by the existence of all
other points. In our DQC analysis this may be replaced in a
straightforward manner by defining dynamic distances among points
\be
d_{i,j}(t)=|| \vev{\vec{x}_i(t)}-\vev{\vec{x}_j(t)} ||
\ee
with the norm being Euclidean or any other suitable choice.

Clearly $d_{i,j}(0)$ is the geometric distance as given by the
original data-matrix or by its reduced form that is being
investigated. As DQC evolves with time $d_{i,j}(t)$ changes, and
when some semi-perfect clustering is obtained, it will be close to
zero for points that belong to the same cluster. Figure~\ref{new2}
shows this change in time for all $d_{i,1}(t)$ in the crab-data
example studied above. It is quite obvious that, in addition to the
few cases in which clustering disagrees with classification, there
are many intermediate steps where different data-points are close to
each other in spite of eventually evolving into different clusters
and belonging to different classes. Thus a close scrutiny of the
dynamic distances matrix $d_{i,j}(t)$ may lead to interesting
observations regarding the relationships among individual pairs of
points in the original data, a relationship that is brought out by
DQC as result of the existing information about all other
data-points. It may be used to further investigate the reason for
such proximities, along any one of the lines mentioned above, and
thus may lead to novel insights regarding the problem at hand.

\section*{Analysis of Large Data Sets}

There are many scientific and commercial fields, such as cosmology,
epidemiology, and risk-management, where the data sets of interest
contain many points, often also in large numbers of dimensions. We
have already discussed how to overcome the problem of large
dimensions. Dealing with large number of points requires yet a new
angle. In problems of this sort it is clear from the outset that,
especially on a PC, diagonalizing matrices which are larger than
$2000 \times 2000$ is computationally intensive. It is obvious that
using brute force methods to evolve sets of data having tens of
thousands of points simply won't work.  The solution to the problem
of dealing with sets of data containing tens of thousands of entries
each with $N$ features, lies in the fact that the SVD decomposition
maps the data into an $N$-dimensional cube, and the fact that the
data points are represented by states in Hilbert space rather than
$N$-tuples of real numbers.  Since there is a literature on ways to
do SVD decomposition for large sets of data, we will not address
this point here.  What we do wish to discuss is how to exploit the
representation of data points as states in Hilbert space in order to
evolve large sets of data.

The trick is to observe that since Gaussian wavefunctions whose
centers lie within a given cube have non-vanishing overlaps, as one
chooses more and more wavefunctions one eventually arrives at a
situation where the states become what we will refer to as {\it
essentially linearly dependent\/}.  In other words, we arrive at a
stage at which any new wavefunction added to the set can, to some
predetermined accuracy, be expressed as a linear combination of the
wavefunctions we already have.  Of course, since quantum mechanical
time evolution is a linear process, this means that this additional
state can be evolved by expressing it as a linear combination of the
previously selected states and evolving them.  Since computing the
overlap of two Gaussians is done analytically(Appendix B)
determining which points determine the set of {\it maximally
essentially linearly independent states for the problem\/} is easy.
Typically, even for data sets with $35,000$ points, this is of the order of
$1000$ points. This works because, as we have already noted, we
don't need high accuracy for DQC evolution. The quality of the
clustering degrades very slowly with loss in accuracy. Thus, we
can compute the time evolution operator in terms of a well chosen
subset of the data and then apply it to the whole set of points.
\endnote{This is particularly attractive for new multi-core PCs and for
computer clusters, since it is possible, even in Maple, to write
multi-threaded programs which farm the multiplication of large
numbers of vectors out to different processors. This means that one
can achieve a great improvement in the speed of the computation for
very little additional work.}

To demonstrate this versatility of DQS we have analyzed a set of
35,213 points in 20 dimensions \endnote{We are grateful to JoAnne
Hewett and Tom Rizzo for providing us with this example.  The points
in the plot represent sample super-symmetric models for physics
beyond the Standard Model that satisfy all experimental constraints in a parameter
space of twenty dimensions.}
This data-set definitely shows some non-trivial variations in
density that can be made apparent by a visual inspection of the data
plotted in different dimensional combinations. However, DQC is
needed to obtain good visual deciphering of the different
structures. Selecting a sub-set of 1200 points, whose Gaussians we
consider to be a set of essentially linearly independent states for
the problem, we construct ${\cal H}^{tr}$.  By expanding
the remaining states in terms of these 1200 states we can
easily evaluate the DQC evolution of all 35,213 points. The
results are displayed in Figure 4. It seems very clear how the
structures develop with DQC. Using the last DQC stage it is possible
to identify the main structures, and assign each substructure a
different color.  One can then examine the colored version of a plot
of the individual data points, discern the structures that belong
together, and follow the DQC devlopment tracing out dynamics
distances between the different points and structures in all
dimensions.

\begin{figure}[h!]
\vbox{
   \hbox to \hsize{\hss
   \includegraphics[width=2.25in]{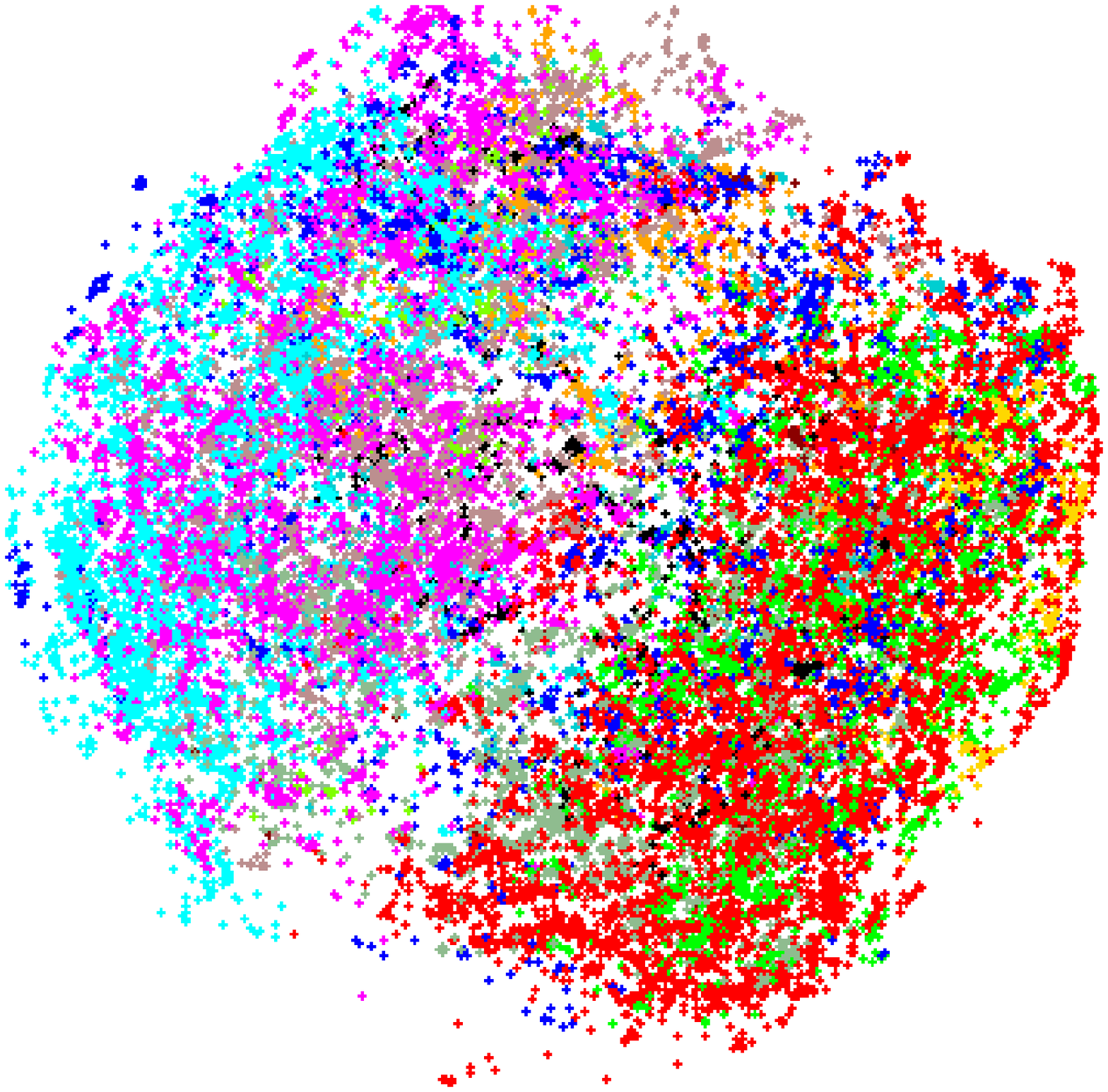}
  \hss\qquad\hss
   \includegraphics[width=2.25in]{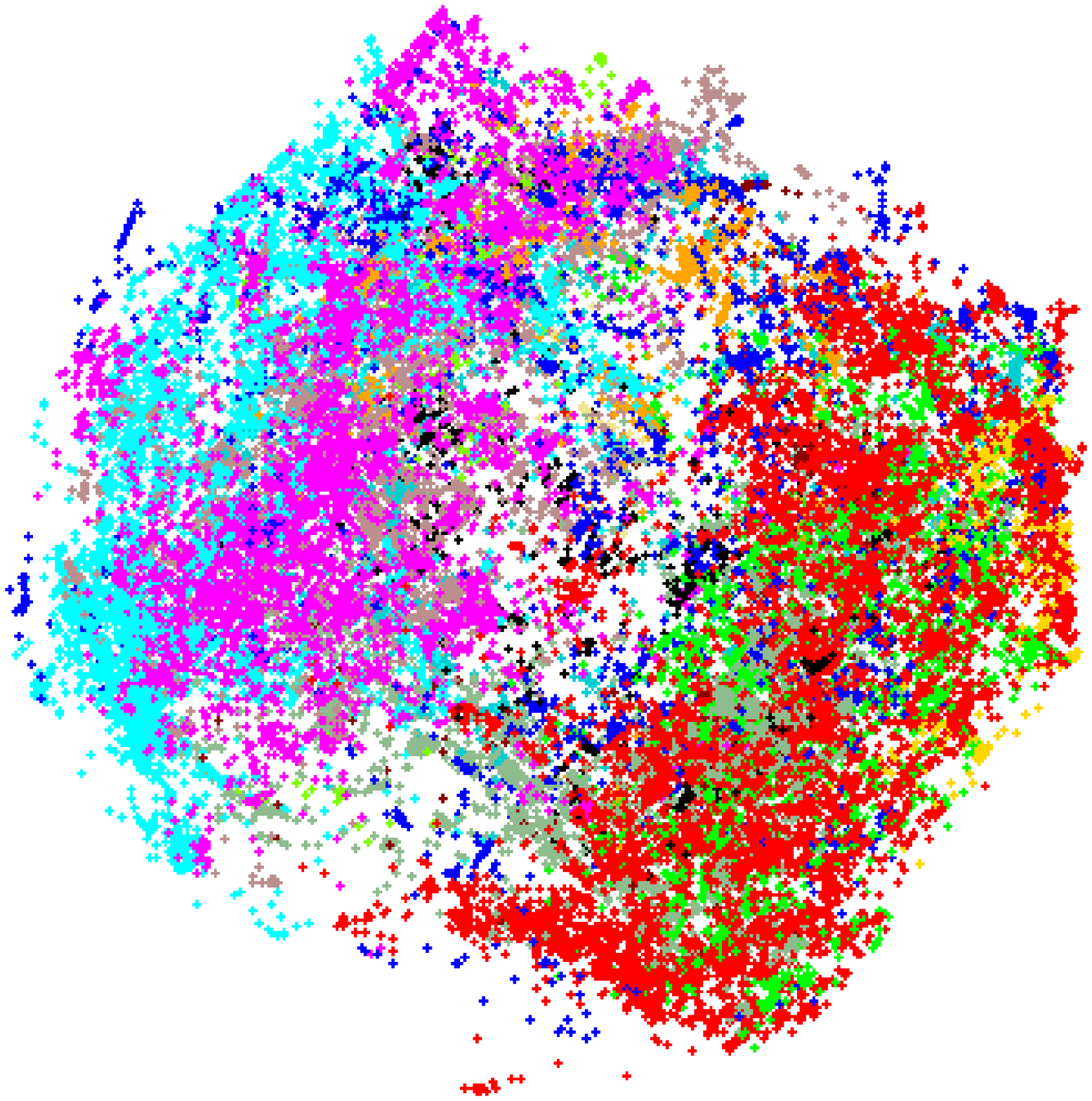}
   \hss\qquad\hss}
   \hbox to \hsize{\hss
   \includegraphics[width=2.25in]{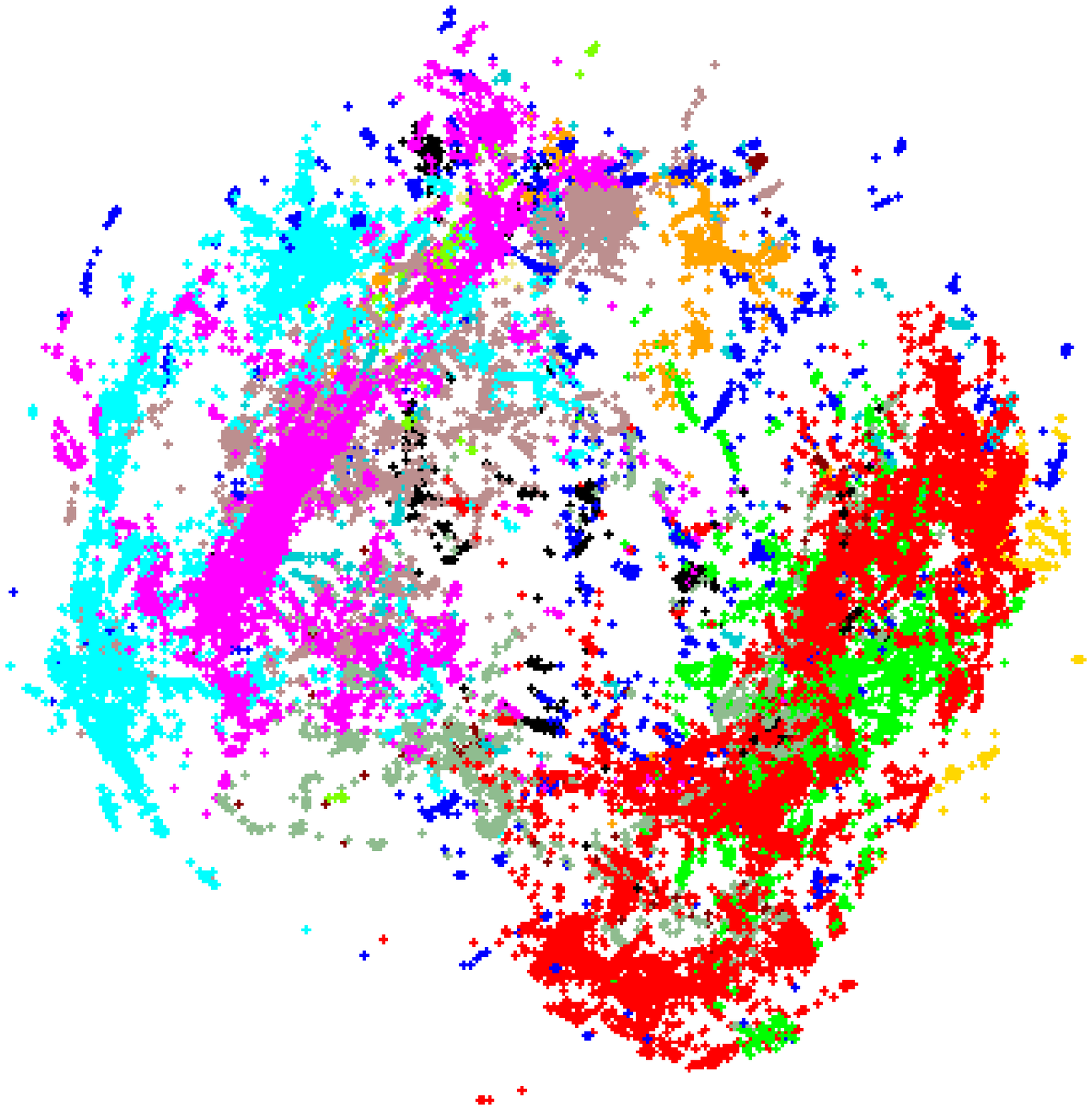}
   \hss\qquad\hss
   \includegraphics[width=2.25in]{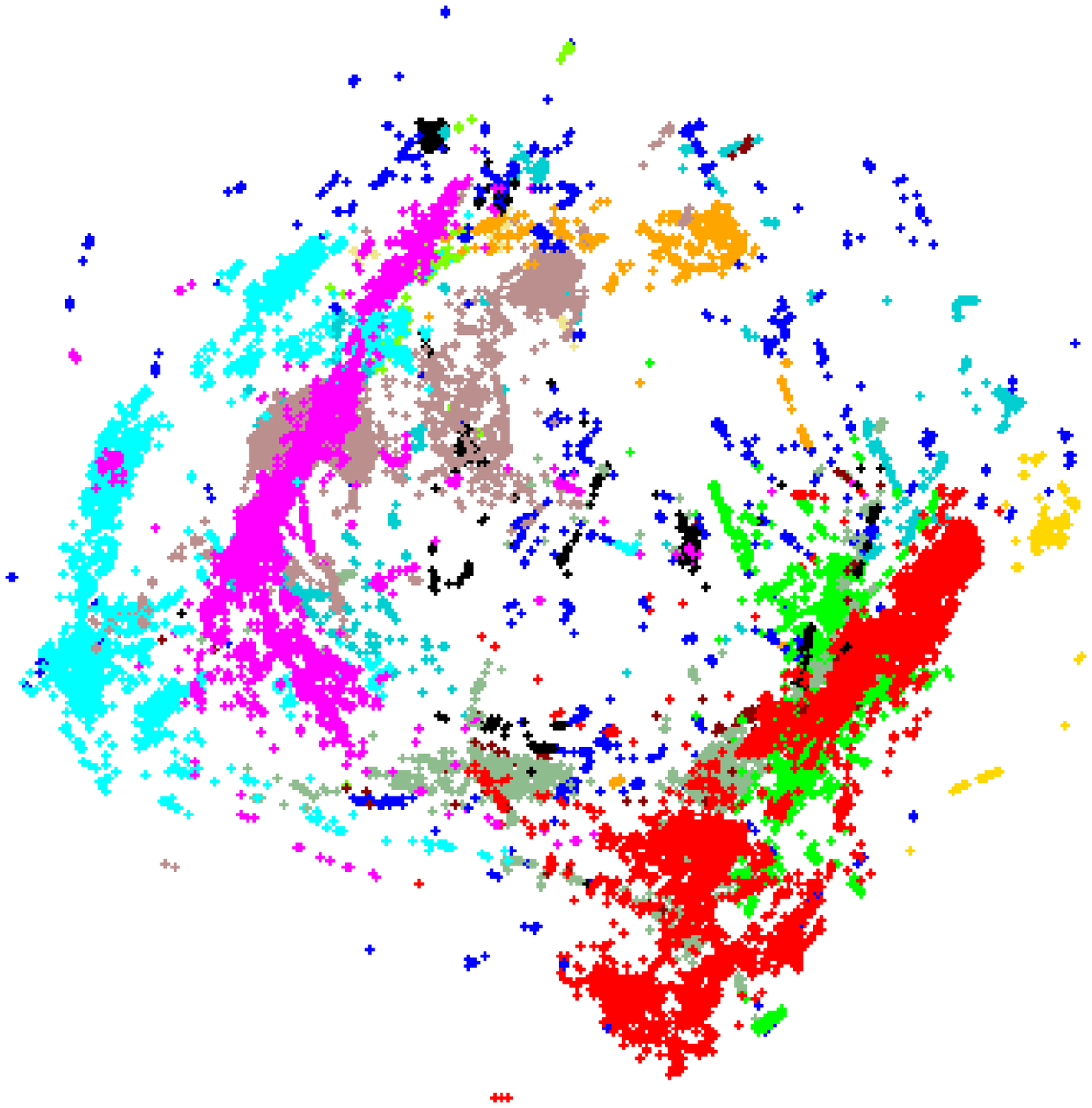}
    \hss}
    }
\caption{A plot of the first three principal components for a large
data-set, comprising 35,213 points in 20 dimensions, before and
after DQC evolution. The potential was determined from the full
data-set and evolved using a sub-set of 1200 points, whose Gaussians
serve as an essentially linearly set of independent states. Three
stages of DQC development are shown. The coloring was decided upon
by selecting the most obvious clusters from the evolved data and
assigning colors to them.  The dark blue points correspond to points
that we did not bother to assign to clusters.  The purpose of
coloring is to be able to look at the points in the original data,
discern those that belong to common structures, and follow their
dynamic distances under DQC evolution.} \label{jandtplot}
\end{figure}

\section*{Interplay of Feature Selection with DQC}

Data exploration involves not only the instances, or data-points,
but also the features (coordinates) with which the instances are
defined. By performing SVD, and selecting a sub-set of coordinates,
we define superpositions of the original features within which we
search for clustering of the instances. In problems with very many
features, it is adventageous to also perform some feature filtering,
employing a judicious selection of subsets of the original features.
Clearly, the effectiveness of preprocessing data using some method
for selecting important features is well appreciated.  What we wish
to show in this discussion is how easily one distinguishes the
effects of feature filtering in our visual approach and how easy it
is, in problems where one has an expert classification, to see if
the unsupervised method used to select important features is working
well.  Furthermore, we wish to show the power of combining
iterations of an SVD based feature filtering algorithm in
conjunction with iterations of DQC.  To do this we will show what
happens when one applies these ideas to the dataset of Golub {\it et
al.} \cite{golub}.

The Golub {\it et.al.} dataset contains gene chip measurements on
cells from 72 leukemia patients with two different types of
Leukemia, ALL and AML. The expert identification of the classes in
this data set is based upon dividing the ALL set into two subsets
corresponding to T-cell and B-cell Leukemia.  The AML set is divided
into patients who underwent treatment and those who did not.  In
total the Affymetrix GeneChip used in this experiment measured the
expression of 7129 genes. The feature filtering method we employ is
based on SVD-entropy, and is a simple modification of a method
introduced by Varshavsky {\it et al.}\cite{featsel} and applied to
the same data.

The method begins by computing the SVD-based entropy \cite{alter} of a dataset $M$
( matrix of $n$ instances by $m$ features of Eq. 15)
based on the eigenvalues $s_j$ of its diagonal matrix $S$. Defining normalized relative
variance values
$v_j = {s_j^2 \over \sum_k s_k^2}$,
the dataset entropy is defined through
\be
E = - {1 \over \log{r} } \sum_{j=1}^r v_j \log(v_j)
\ee
where $r$ is the rank of the data-matrix, typically much smaller than $m$ .
Given the dataset entropy of the matrix $M$, define the contribution of the
$i^{th}$ feature to the entropy using a leave-one-out comparison; i.e.,
for each feature we construct the quantity
\be
    CE_i = E(M_{(n \times m)}) - E(M_{(n \times (m-1))})
\ee
where the second entropy is computed for the matrix with the $i^{th}$
feature removed.  Our filtering technique will be to remove all features
for which $CE_i \le 0$.

\begin{figure}[h!]
   \hbox to \hsize{\hss
   \includegraphics[width=2.75in]{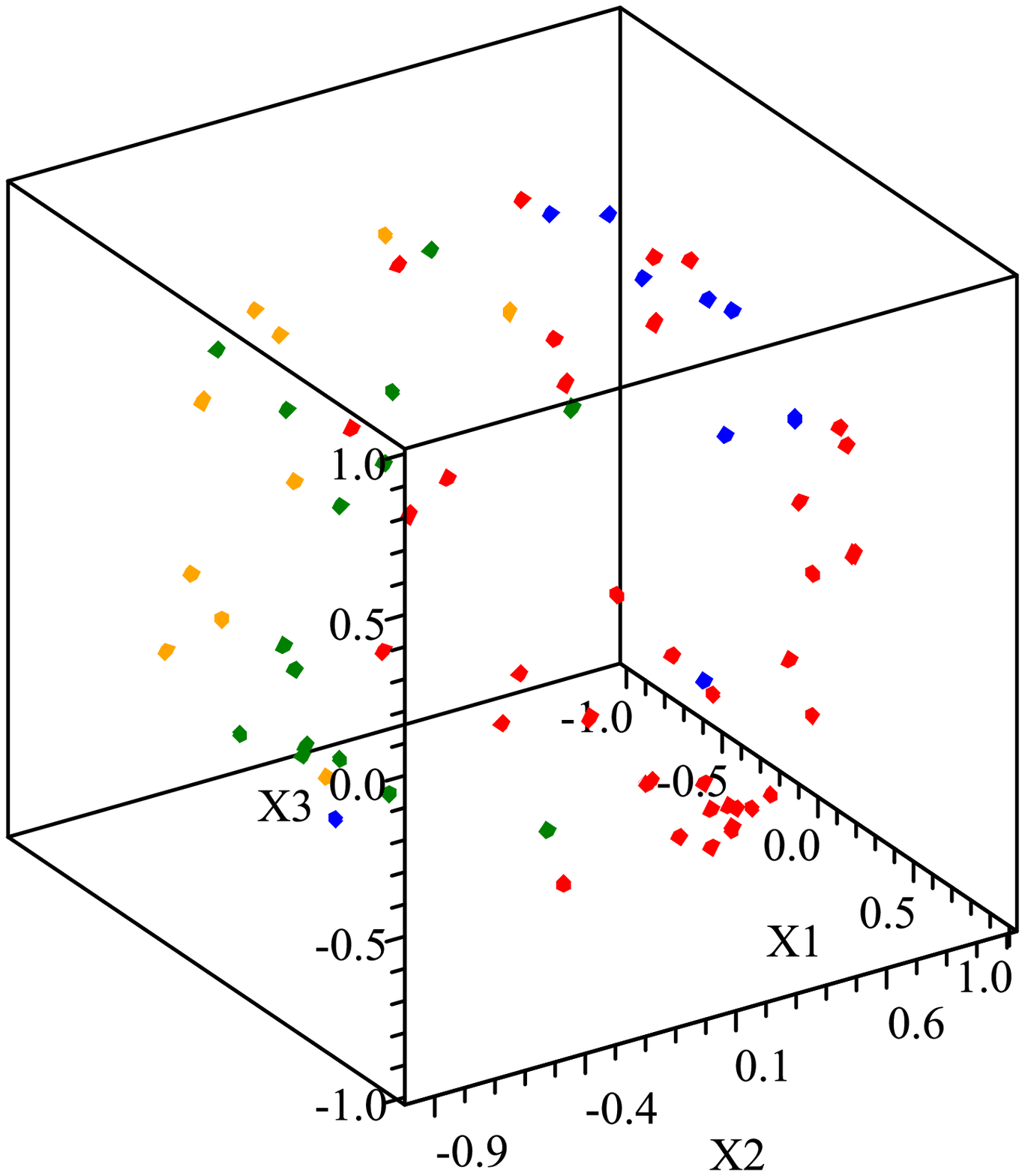}
  \hss\quad\hss
   \includegraphics[width=2.75in]{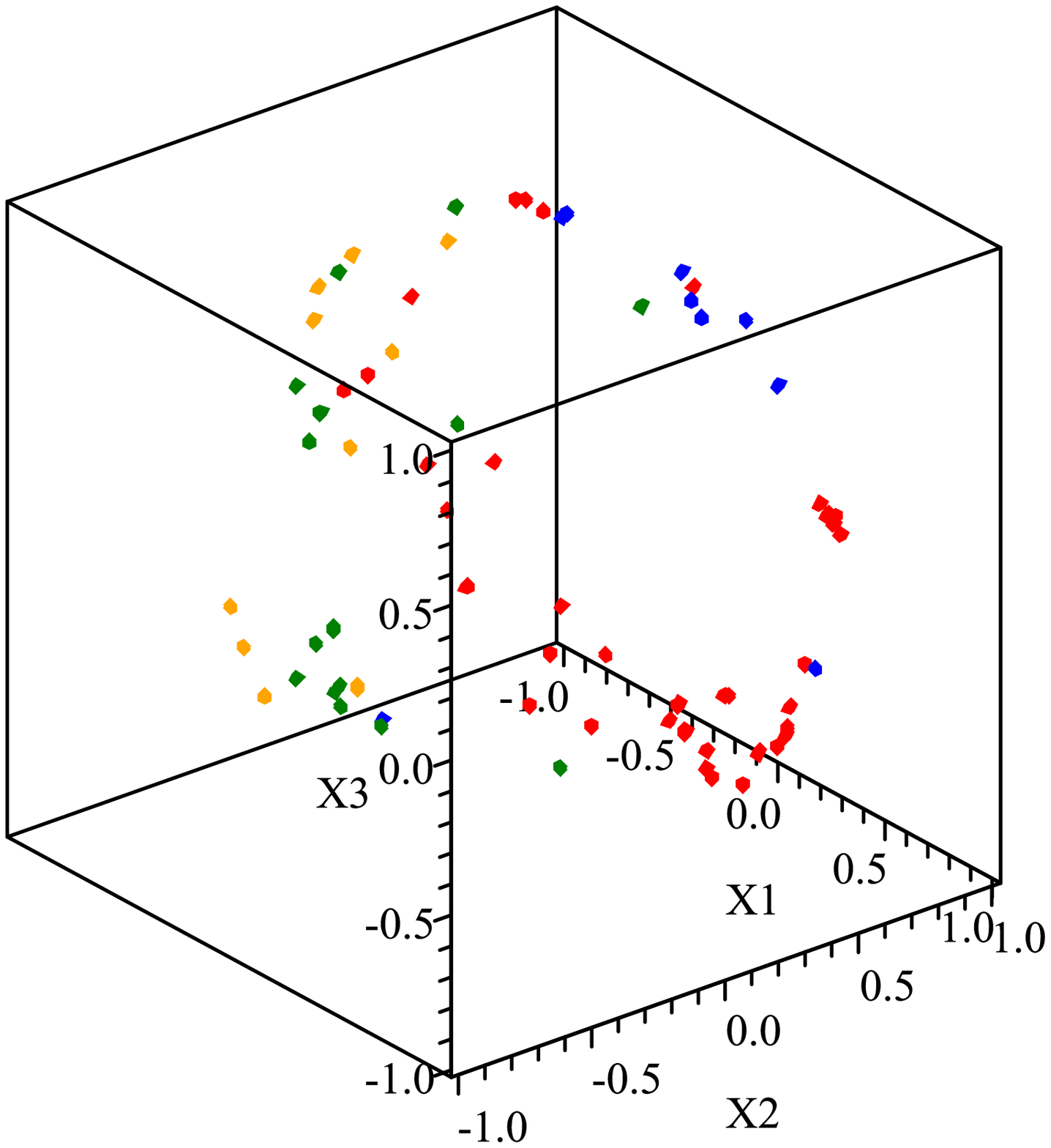}
    \hss}
  \caption{The left hand picture is the raw data from the
  Affymetrix Chip plotted for principal components 2,3,4.
  Clearly, without the coloring it would be hard to identify
  clusters.  The right hand picture is the same data after DQC
  evolution using $\sigma =0.2$ and a mass $m=0.01$.  The different
  classes are shown as blue, red, green and orange.
   }
\label{GolubOne}
\end{figure}

\begin{figure}[h!]
   \hbox to \hsize{\hss
   \includegraphics[width=2.75in]{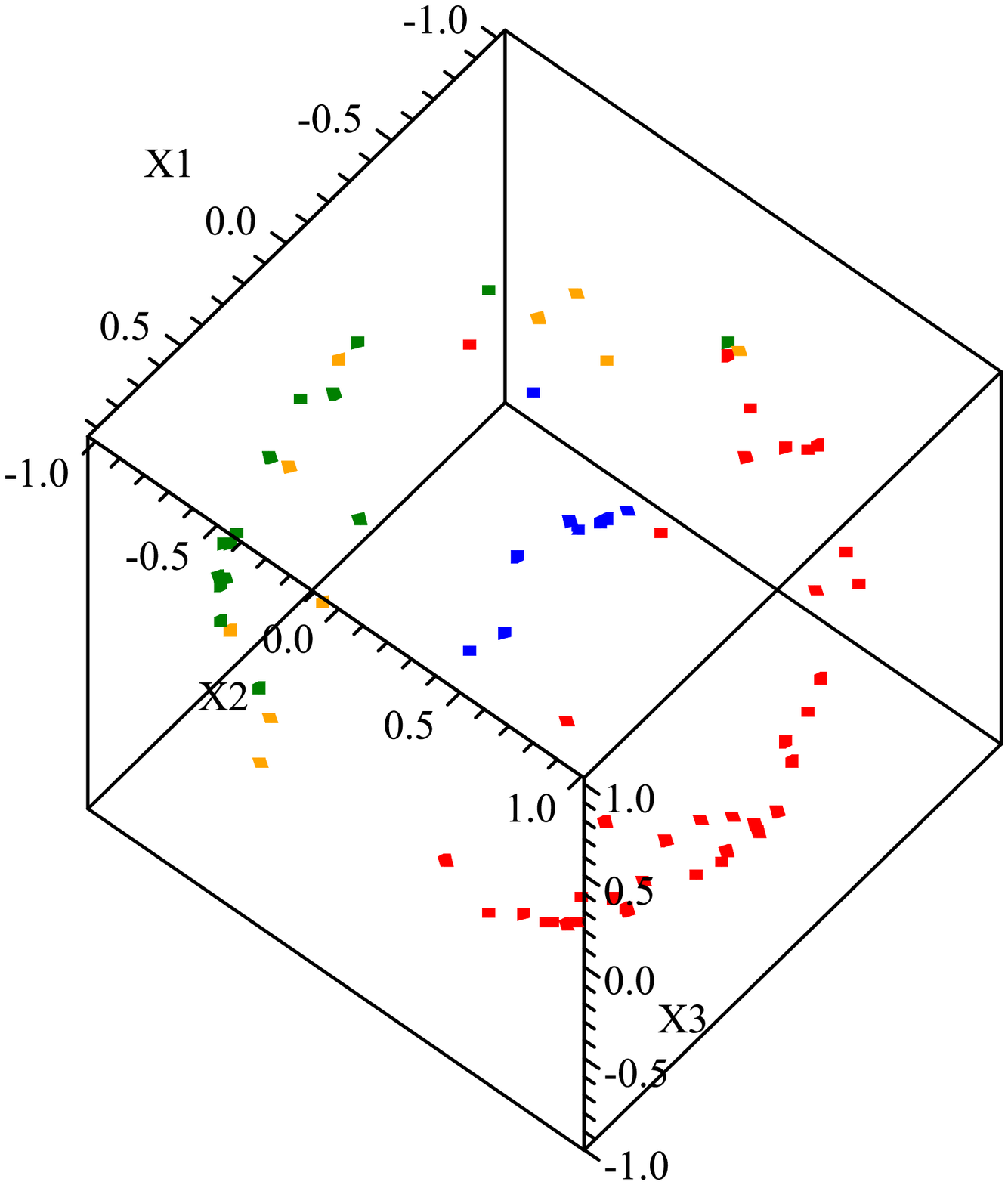}
  \hss\quad\hss
   \includegraphics[width=2.75in]{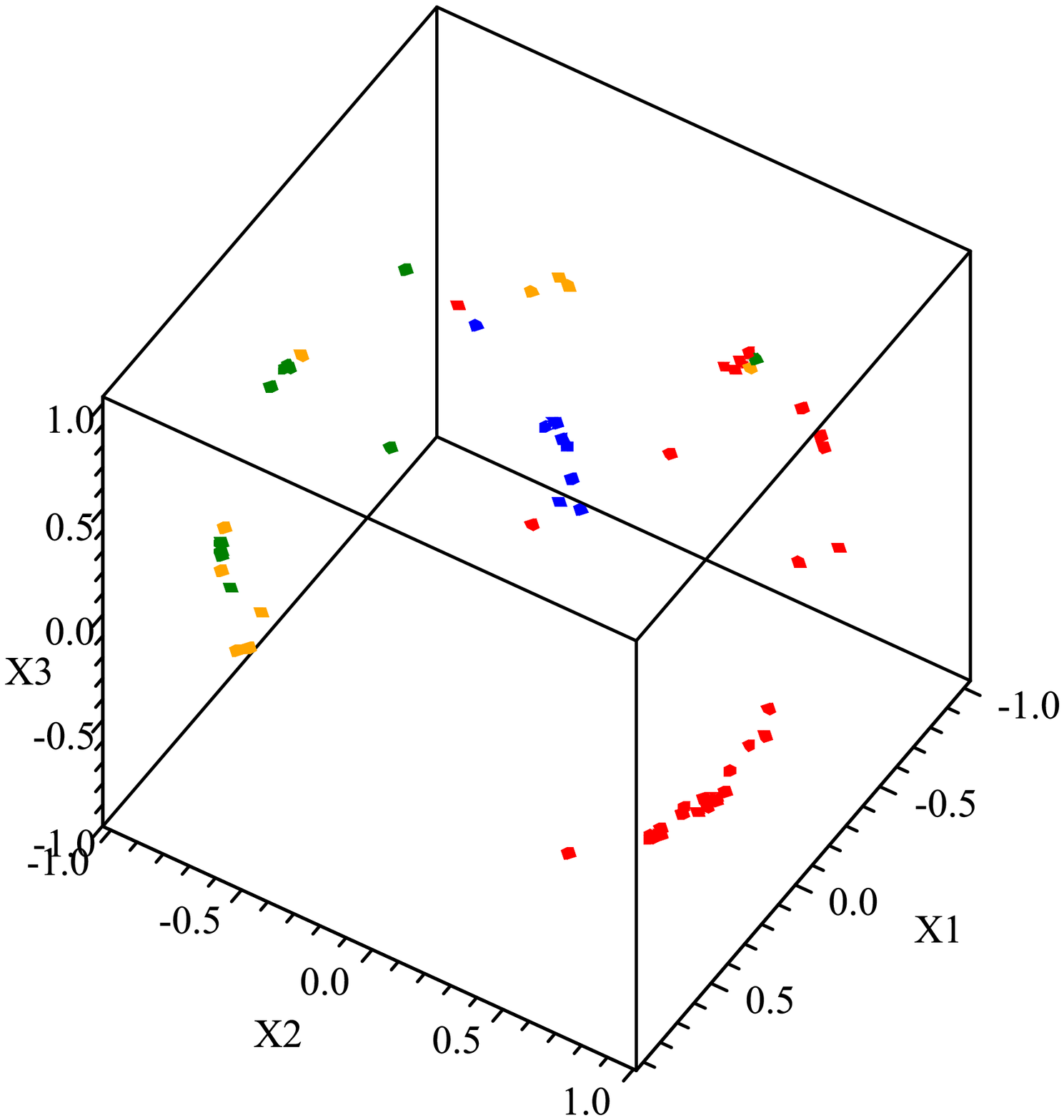}
    \hss}
  \caption{The left hand plot is the Golub data after one stage
  of SVD-entropy based filtering, but before DQC evolution.  The right
  hand plot is the same data after DQC evolution.
   }
\label{GolubTwo}
\end{figure}

Figure~\ref{GolubOne} displays the raw data in the 3-dimensional
space defined by PCs 2 to 4, and the effect that DQC has on these
data. In Figure~\ref{GolubTwo} we see the result of applying feature
filtering to the original data, represented in the same
3-dimensions, followed by DQC evolution.  Applying a single stage of
filtering has a dramatic effect upon clustering, even before DQC
evolution. The latter helps sharpening the cluster separation.

\begin{figure}[h]
   \hbox to \hsize{\hss
   \includegraphics[width=2.75in]{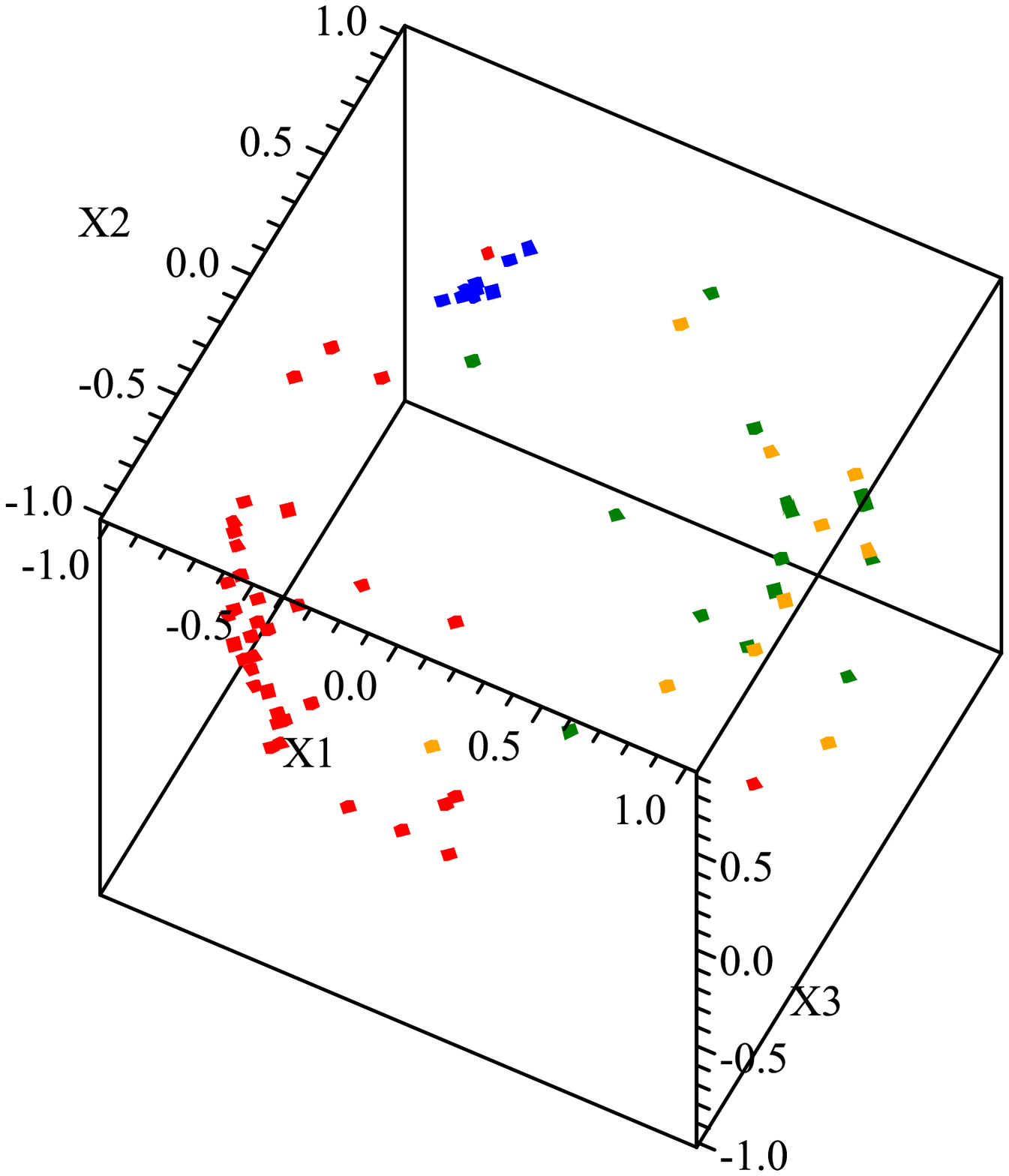}
  \hss\quad\hss
   \includegraphics[width=2.75in]{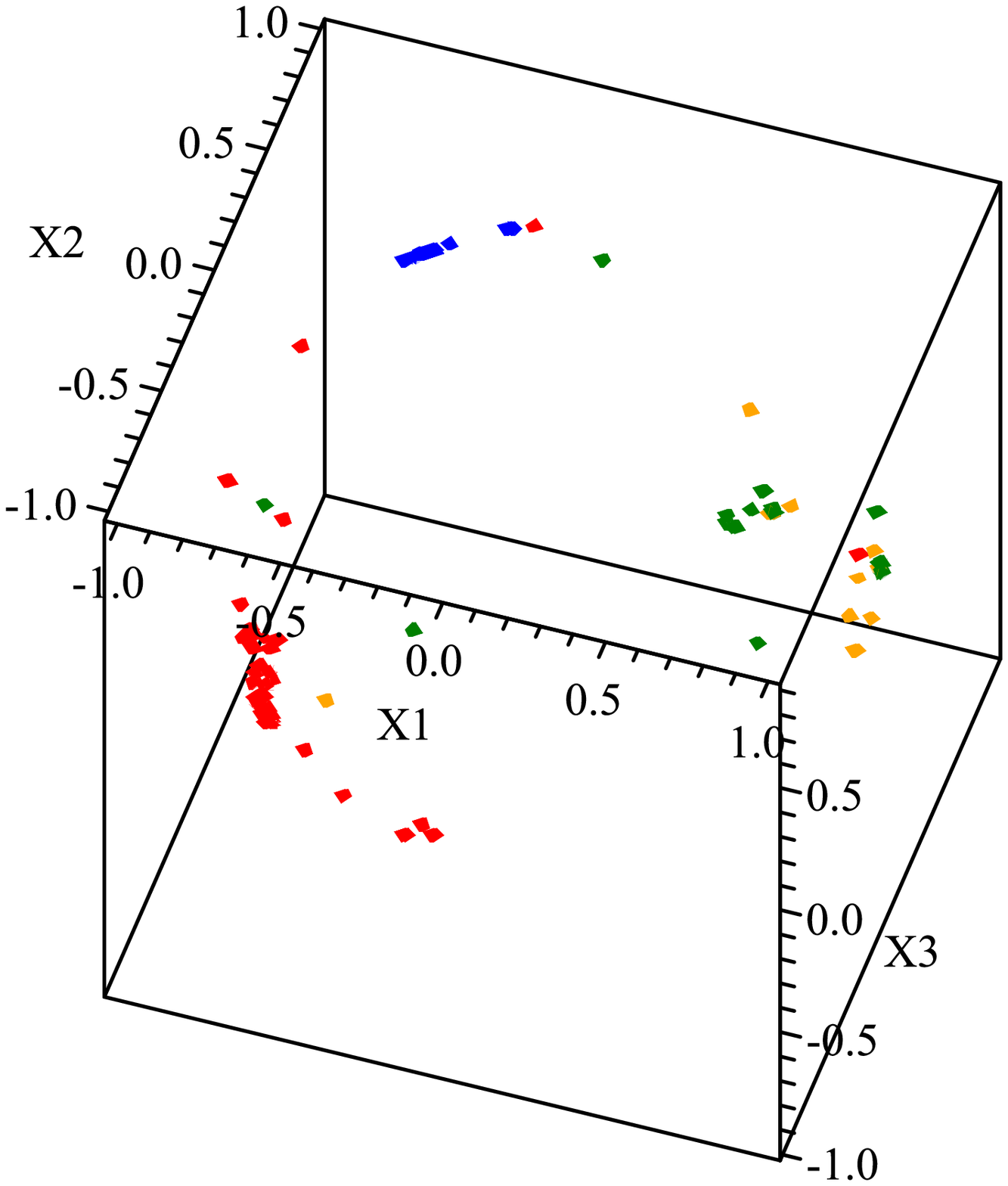}
    \hss}
  \caption{The left hand plot is the data after three stages
  of SVD-entropy based filtering, but before DQC evolution.  The right
  hand plot is the same data after DQC evolution.
   }
\label{GolubThree}
\end{figure}


Figure \ref{GolubThree} shows the results of
three iterations of SVD-entropy, before and after DQC
evolution.  These plots, especially the after DQC pictures, show
dramatic clustering, especially for the blue points.  With each
stage of filtering we see that the blue points cluster better and
better, in that the single red outlier separates from the cluster
and the cluster separates more and more from the other points.  The
blue points are what we will refer to as an {\it obviously robust
cluster\/} which has been identified in early stages of filtering.
If one continues iterating past the fifth stage, however, the clear
separation of the blue points from the others begins to diminish.
Thus we see that the SVD-entropy based filtering, in trying to
enhance the clumping of the red points, starts throwing away those
features which make the blue cluster distinct.  Since this effect is
quite pronounced we would say that features that are important to
distinguishing the blue cluster from the others begin to be removed
at the sixth and higher iterations of filtering.  This is, of
course, just what we are looking for, a way of identifying those
features which are important to the existing biological clustering.
Out of the original 7129 features, we have reduced ourselves to 2766
features by the fifth iteration.  In going from step five to step
six this gets further reduced to 2488 features, so we could begin
searching among the 278 eliminated features to isolate those most
responsible for the separation of the blue cluster from the others.
Instead, we will take another track and, since it is so robust and
easily identified, remove the blue cluster from the original data
and repeat the same process without this cluster.  The idea here is
that now the SVD-entropy based filtering will not be pulled by the
blue cluster and so it will do a better job of sorting out the red,
green and orange clusters.  As we will see, this is in fact the
case.

\begin{figure}[h]
   \hbox to \hsize{\hss
   \includegraphics[width=2.75in]{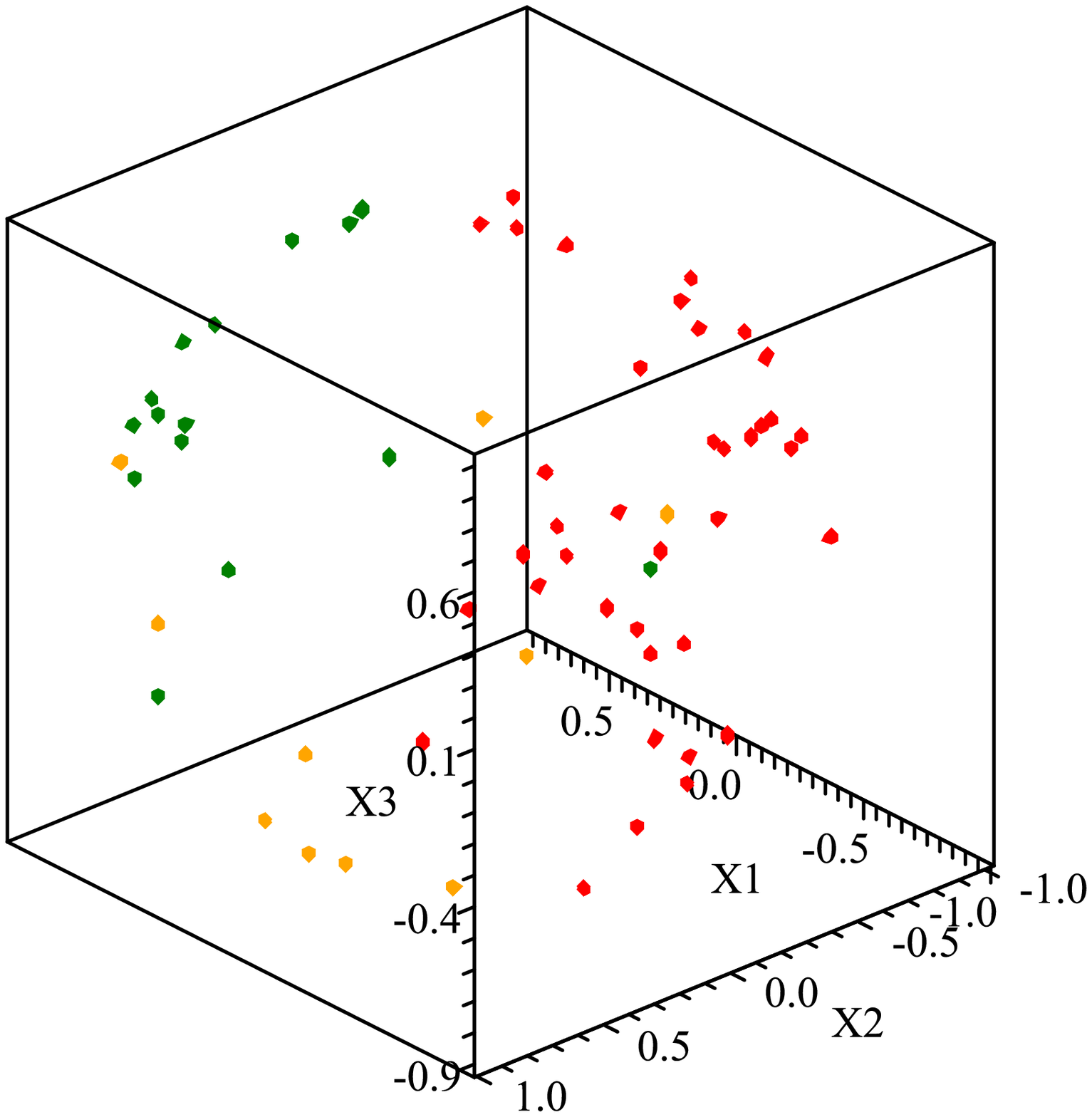}
  \hss\quad\hss
   \includegraphics[width=2.75in]{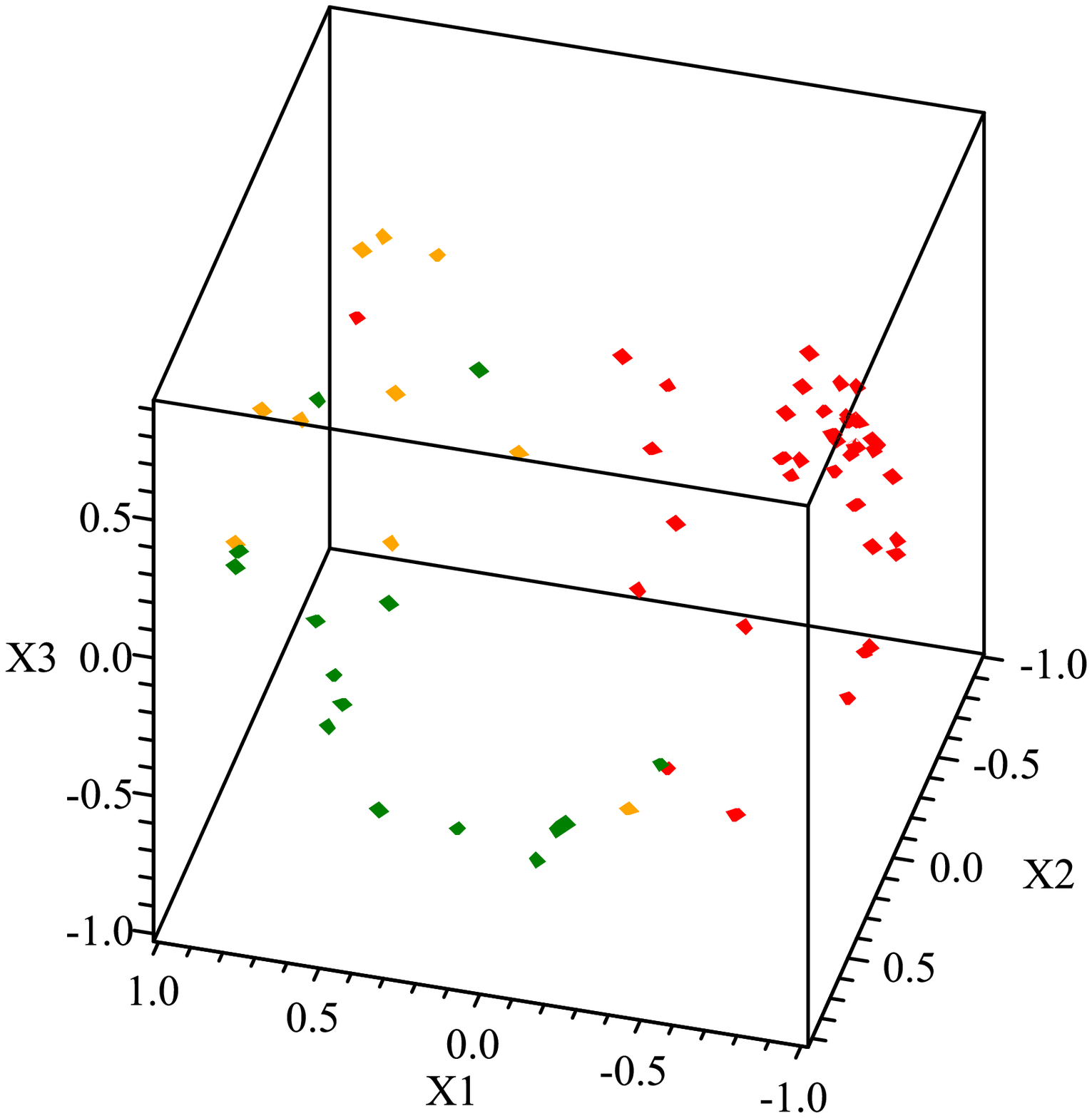}
    \hss}
  \caption{The left hand plot is what the starting data looks like
  if one first removes the blue points and does one stage of SVD-entropy
  based filtering.  The right hand plot is what the starting data looks
  like after three stages of filtering.
   }
\label{GolubFive}
\end{figure}

In Figure \ref{GolubFive} we see a plot of what the starting
configurations look like if one takes the original data, removes the
identified blue cluster and re-sorts the reduced data set according
to the SVD-entropy based filtering rules.  The left hand plot is
what happens if one filters a single time, removing those features,
$i$, whose one-left-out comparison, $CE_i$, is less than or equal to
zero. The right hand plot shows what happens if one repeats this
procedure two more times, each time removing features for which
$CE_i \le 0$. There is no problem seeing that each iteration of the
SVD-entropy based filtering step improves the separation of the
starting clusters.  By the time we have done five SVD-entropy based
filtering steps the red, green and orange clusters are distinct, if not
obviously separated.

\begin{figure}
   \hbox to \hsize{\hss
   \includegraphics[width=2.75in]{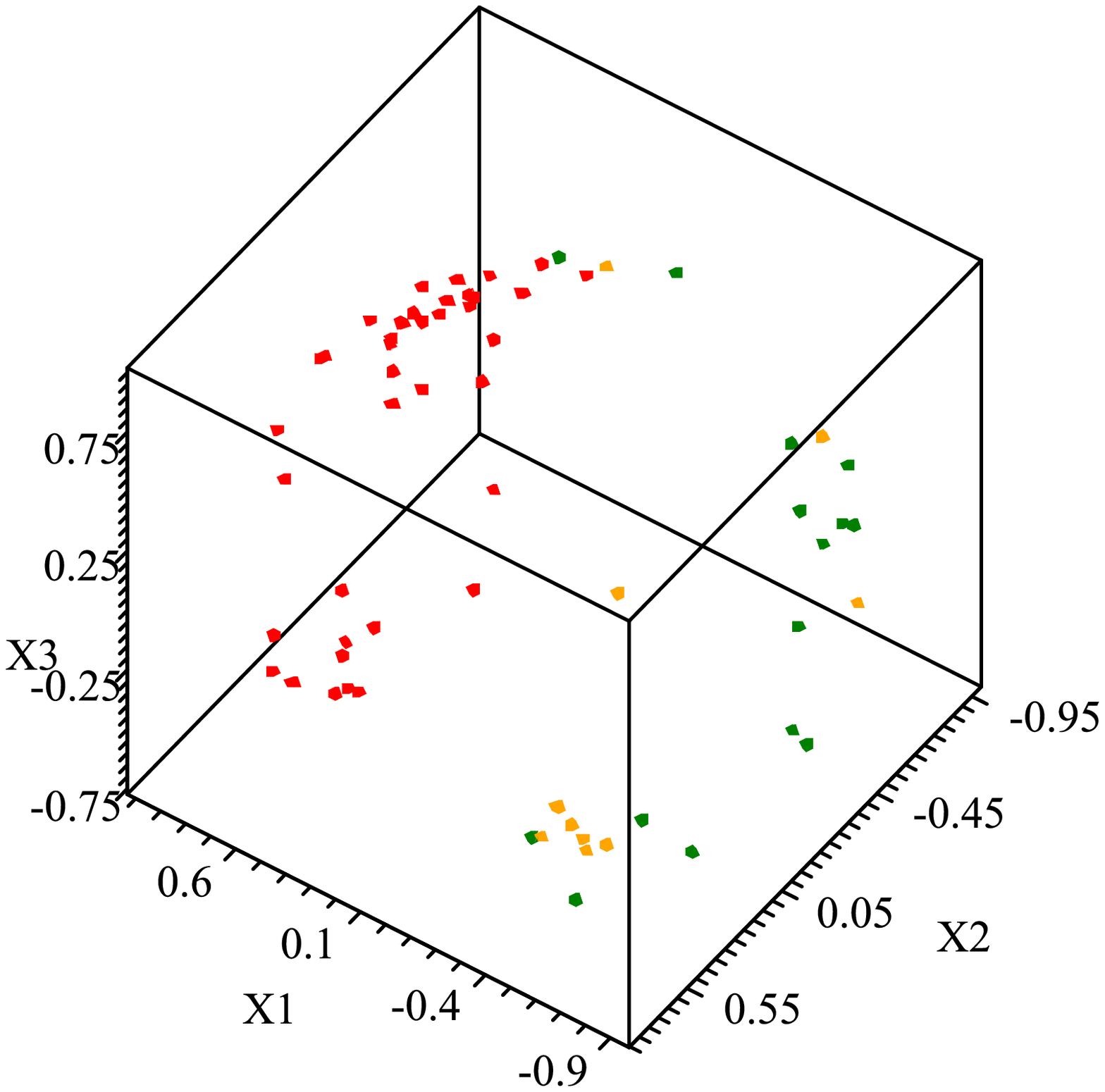}
  \hss\quad\hss
   \includegraphics[width=2.75in]{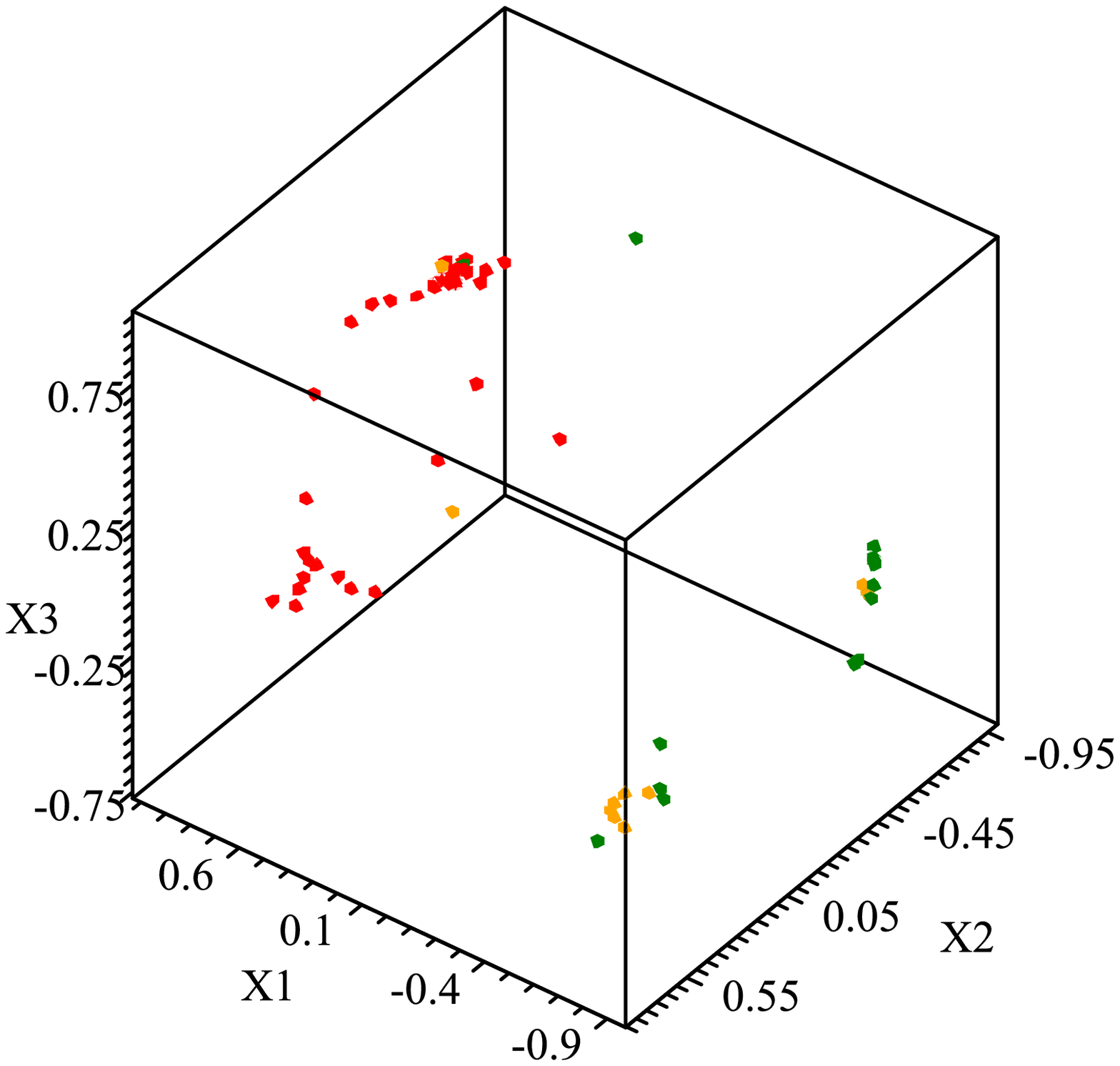}
    \hss}
   \hbox to \hsize{\hss
   \includegraphics[width=2.75in]{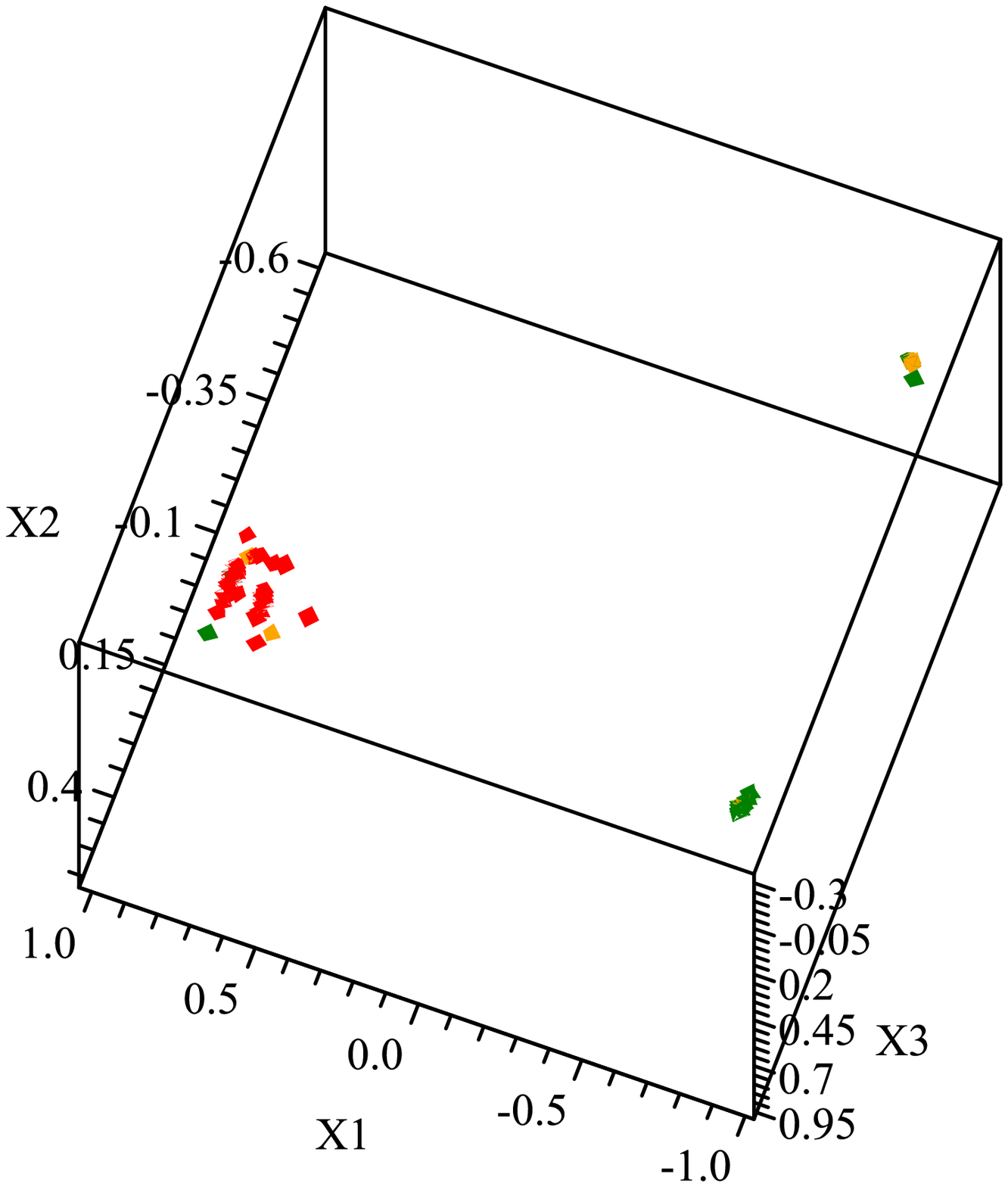}
   \hss}
  \caption{The left hand plot is what the starting data looks like
  if one first removes the blue points and does five stages of SVD-entropy
  based filtering.  The right hand plot is what happens after one stage
  of DQC evolution.  The bottom plot is the final result after iterating
  the DQC evolution step two more times.  At this point the clusters
  are trivially extracted.
   }
\label{GolubSix}
\end{figure}

Finally, to complete our discussion, we show Figure \ref{GolubSix}.
This figure shows the results of doing five iterations of the
SVD-entropy based filtering and following that with three stages of
DQC evolution.  The dramatic clustering accomplished by DQC
evolution makes it easy to extract clusters.  Note however, that in
the second plot we see what we have seen throughout, that the red
points first form two distinct sub-clusters which only merge after
two more stages of DQC evolution.  This constant repetition of the
same phenomenon, which is only made more apparent by SVD-entropy based
filtering, is certainly a real feature of the data.  It presumably
says that what appears to be a sample of a single type of cell at
the biological level is in reality two somewhat different types of
cells when one looks at gene expression. A measure of the success of clustering
is given by the Jaccard score \endnote{ The Jaccard score is
evaluated by considering all pairs of data points, and asking if
they cluster together and if they fit in the same class, as judged
by the expert. The Jaccard score is then defined by $J= {tp \over tp
+ fp + fn}$ where $tp,\,fp,\,fn,$ stand for true-positive,
false-positive and false-negative, correspondingly.} which, for this result
is $0.762$, higher than the value 0.707 obtained by \cite{featsel}.

\section*{Summary and Outlook}

We have proposed a dynamical method for exploring proximity
relationships among data-points in large spaces.  Starting with the
potential function of quantum clustering \cite{qc1} we have shown
how to embed it into a dynamical theory so as to provide a visual
exploratory tool.  Formulating the theoretical treatment using coherent
(Gaussian) states allows us to derive analytic expressions for all
necessary calculations of the temporal evolution.  This allows us to
treat quite complicated data and put them into a visual framework
that can be easily manipulated by the user who wishes to search for
structures in the data.  We have tested the system on random data to
make sure that it does not produce unwarranted clustering
structures.

Throughout this paper we represent the DQC evolution of the Gaussians
associated with the original data-points, by following the centers
of the evolving wave-functions. It should be noted that there is more
information to be gained from the full wave-function of a data-point: it is expected
to expand, at times, and cover a large fraction of the domain of the cluster
with which this point is associated. It may also tunnel into neighboring
clusters with which the point has small dynamic distances.
We expect this notion to be particularly useful when the data may be
better described in terms of 'elongated clusters', i.e. when cluster cores
are not points but lines (e.g. a ring) or higher-dimensional manifolds.
Note that our methodology is not confined to potentials that have only
well-separated minima.

We have discussed the virtues of combining DQC with some
preprocessing tools. The first was SVD, which was used to limit the
range of the data values and to allow us to do some dimensional
reduction.  While dimensional reduction is a must for handling data
in very large dimensions, and it helps to remove noise from the
data, we wish to point out that DQC can handle a large number of
features without much difficulty.  The computational complexity of
the problem is controlled by the number of data-points, since this
defines the size of the matrix to be exponentiated. The
computational cost associated with keeping more features is only
related to computing the matrices associated with multiplying a
wave-function by a given coordinate.  This is a one time cost.
The computational cost of computing the values of these operators
only grows linearly with the number of features. Clearly it is
possible to avoid these costs by keeping a large number of features
when constructing the quantum potential, $V(\x)$, and plotting a
much smaller number of features when constructing the animations.
Experience has shown that after one stage of DQC evolution,
clustering which occurs because of structures in $V(\x)$ that are
only seen in features that are not plotted in the animations becomes
readily visible in those plots that we do construct.  This aspect of
DQC allows us to avoid some of the problems associated with using
SVD to strongly reduce the number of dimensions. In addition to
dimensional reduction based upon simply making an SVD decomposition
of the data, we discussed one scheme for selecting individual
features that are judged to be relevant to the data at hand.  Since
our problem is unsupervised, we employed a feature filtering method
that depends on the contribution of the features to SVD-entropy.
The examples showed that the visual nature of DQC made it easy
to judge the effectiveness of feature filtering, especially after
iterative applications of DQC evolution.

We have already noted, that for sets of data containing entries with
a very large number of features, DQC has the computational advantage
that once one has formed the Hamiltonian of the system, the
computational problem is carried out using a matrix which has no
more rows and columns than the number of data points. Moreover, we
have seen that the simplest reduction of the analytic problem of
assigning data points to minima of the multi-dimensional potential
function works remarkably well. Going beyond the truncation
procedure explained in Appendix B, while easily doable, seems
unnecessary for most problems, and this allows us to greatly speed
up the computations. In our analysis we went on to discuss the case of
data sets containing large numbers of points. It turns out that, using
our Hilbert space representation of data-points, we can naturally select
a small set of points whose Gaussians span efficiently the entire
Hilbert space. These Gaussian are then used as the basis for calculating the DQC
evolvement of ${\bf all}$ points. It is quite obvious from the
example displayed in Fig. 4 how well these properties of DQC can be employed
to discern structures in the large data-set under consideration.

Finally, we wish to observe that the DQC methods described in this
paper can be easily extended to general classification problems that
are usually resolved by supervised machine learning methods.  The
point is that given a training set, i.e., a data set that has been
fully resolved by DQC once the appropriate stages of dimensional
reduction and feature filtering has been applied, then one can use
this set to classify new data.  There are two different ways one
can accomplish this task.  In the first approach we use the fact
that the training set has been successfully clustered to assign distinct
colors to points that lie in the training set, so that they may be
visually identify in all subsequent studies.  Once this has been done, the
classification of new data points can been accomplished in two
steps. First, reduce the SVD matrix containing both the training set
and the new data points (using the previously determined features)
to an appropriate dimension, and construct the QC potential
for the full data set including the training set. Next, apply DQC to study the
evolution of the full system using the new QC potential and see how
the new points associate themselves with the points in the training
set.   Note, as always, both the intermediate dynamics and eventual
coalescence of the full set into clusters can give useful
information about the full data set.  The fact that the old points
have been colored according to the original classification scheme
makes it possible to see if the SVD reduction of the full data set
(training set plus new data) distorts the original classification.
If this happens, i.e. if the original points fail to cluster
properly, then one can go back and use the tools of feature
filtering, etc. to analyze what has changed.  This sort of visual
identification of aspects of the data which distort clustering was
already used in the case of the leukemia data set to see that the
existence of a {\it strong\/} cluster can distort the clustering of
the remaining data.  Once this easily identified cluster was removed
from the data set the clustering of the remaining data was
significantly improved.

The second approach, which is necessary if the dataset contains many
entries and the training set is itself large, is to use only the
training set to generate the quantum potential and the exponential
of the Hamiltonian.  Next, as we already discussed, use this operator
to evolve the full dataset, including the training set.  In this
case the training set is guaranteed to cluster as before and we
can categorize the new points according to how they associate
with known clusters in the training data.


\section*{APPENDIX A. USEFUL OPERATOR IDENTITES}

Using conventional quantum-mechanical notation we represent the Gaussian wave function by
\be
    \ket{\sigma}=({{\sqrt{\pi}}\sigma})^{-{\half}}\,e^{-x^2/2 \sigma^2},
\ee
where we adopted Dirac's bra and ket notation \cite{qm} to denote
$\ket{\psi} = \psi(x) $ and
$\bra{\psi} = {\psi}(x)^\ast$.
Employing the operators $x$ and $ p = {1 \over i} { d \over dx}$ obeying the
commutation relations $\left[ x, p \right] = i $, we define the annihilation operator
\be
    \A = i\,{\sigma \over \sqrt{2}}\,p + {1 \over \sigma \,\sqrt{2}}\, x
\ee
obeying
$
   \A \ket{\sigma} = 0.
$
Its Hermitian adjoint creation operator
$
    \Adag = -i\,{\sigma \over \sqrt{2}}\,p + {1 \over \sigma \,\sqrt{2}}\, x
$
obeys
$
    \left[ \A, \Adag \right] = 1 .
$

We will need a few identities to derive the matrix elements we have to calculate.
First we note the normal ordering identity (meaning
rewriting by using the operator commutation relations so that $\A 's$ appear to
the right of all $\Adag 's$):
\be
    e^{\alpha (\Adag + \A)} = e^{\alpha^2/2}\,e^{\alpha \Adag} \,e^{\alpha \A}
\label{normalorderx}
\ee
which may be proven by differentiation with respect to $\alpha$.
Next we note that
\be
e^{g(\alpha) \Adag}\,\A\,e^{-g(\alpha)\Adag} = \sum_n \frac{g(\alpha)^n}{n!} [ \Adag,[\Adag,[
\ldots,[\Adag,\A]]]\ldots ]_n = \A - g(\alpha)
\label{multiplecomm}
\ee
which is easily derived by differentiating with respect to $g$ and noting that only the first
commutator is non-zero.
A similar calculation proves the equally useful result:
\be
 e^{\alpha ( \Adag - \A )} = e^{-\alpha^2/2}\,e^{\alpha \Adag} \,e^{-\alpha \A}
\label{normalorderp}
\ee

Now, because the Parzen window estimator is constructed using Gaussian wavefunctions
centered about points other than $x=0$, it is convenient to have an operator expression
which relates the Gaussian centered about $x=0$ to the Gaussian centered about
$x = \bar{x}$.

{\bf Theorem:}
 $\ket{\sigma,\bar{x}} = e^{-i p \bar{x}}\,\ket{\sigma} $
is a normalized Gaussian wave-function centered at $x = \bar{x}$; i.e.
\be
\ket{\sigma,\bar{x}} = ({{\sqrt{\pi}}\sigma})^{-{\half}}\,e^{-{(x-\bar{x})^2 \over 2 \sigma^2}}.
\ee
This state is known as a coherent state \cite{klauder}, obeying
\be
\A \ket{\sigma,\bar{x}} = \bar{x} \ket{\sigma,\bar{x}}.
\ee
The generalization to Gaussians in any number of dimensions is straightforward, since they are just
products of Gaussians defined in each one of the different dimensions.

\section*{APPENDIX B. MATRIX ELEMENTS}

The states we start out with $\ket{\sigma,{\bar{x}}_i}$ have norm one and
are, in general, linearly independent; however, they are not orthogonal to
one another.   In what follows we will need an explicit formula
for the scalar product of any such Gaussian $\ket{\sigma,{\bar{x}}_i}$
with another $\ket{\sigma,{\bar{x}}_j}$.  This is easily derived
given the operator form for the shifted Gaussian derived in Appendix~A.
Thus we find that
\be
    \bracket{\sigma,\bar{y}}{\sigma,\bar{x}} = \bra{\sigma}\,e^{-ip( \bar{x} - \bar{y})}
    \ket{\sigma}= e^{-(\bar{x}-\bar{y})^2/4 \sigma^2},
\ee
which is needed for computing the matrix of scalar products
$
    N_{ij} = \bracket{\sigma,\bar{x}_i}{\sigma,\bar{x}_j}.
$
Similarly, by employing
$
e^{i p \bar{y}}\,x\,e^{-i p \bar{y}} = x + \bar{y}
$
we find that
\be
 \bra{\sigma,\bar{y}}\, x\, \ket{\sigma,\bar{x}} =
{(\bar{x}+\bar{y}) \over 2}\,e^{-(\bar{x}-\bar{y})^2/4 \sigma^2}.
\ee
It is straightforward to generalize this derivation to obtain
\be
 \bra{\sigma,\bar{y}}\,V(x)\, \ket{\sigma,\bar{x}} =
e^{-(\bar{x}-\bar{y})^2/4 \sigma^2} \bra{\sigma}\,
V(x + {(\bar{x}+\bar{y}) \over 2})\,\,\ket{\sigma},
\ee
for any function $V(x)$.
Note that this expectation value
can be evaluated by expanding $V$
in a Taylor series about the point $ (\bar{x}+\bar{y})/2$. The leading term is
simply
$
e^{-(\bar{x}-\bar{y})^2/4 \sigma^2}\,V\left({\bar{x}+\bar{y} \over 2}\right)
$
and the remaining terms, involving $\bra{\sigma}\,x^n\,\ket{\sigma}$
can be evaluated from the identity
\be
    \bra{\sigma}\, e^{\alpha\, x} \,\ket{\sigma} =
    \sum_{n=0}^{\infty} {\alpha^n \over n!} \bra{\sigma} x^n \ket{\sigma}
    = \sum_{p=0
    }^{\infty} {\alpha^{2p} \sigma^{2p} \over 4^p\,p!}.
\ee

To speed up computations
we chose to approximate all expectation values of $V(x)$ by $V({\bar{x}+\bar{y} \over 2})$,
the first term in this series.
One could obviously get a more accurate approximation to
the original problem by including additional terms but explicit computation
has shown that, for our purposes, this level of accuracy is sufficient.

The final formula we need to derive is that for
\be
    \bra{\sigma,\bar{y}} \, p^2 \, \ket{\sigma,\bar{x}} =
    \bra{\sigma}\, p^2 \,e^{-i p (\bar{x}-\bar{y})}\, \ket{\sigma}=
    {(\bar{x}-\bar{y})^2 \over 2 \sigma^2} \,e^{-(\bar{x}-\bar{y})^2/4 \sigma^2}.
\ee

With these preliminaries behind us it only remains to describe the mechanics
of the DQC evolution process, where we evaluate the
Hamiltonian truncated to an $n \times n$ matrix in the non-orthonormal
basis of shifted Gaussians:
\be
   {\cal H}_{i,j} = \bra{\sigma,\bar{x_i}}\, H \, \ket{\sigma,\bar{x_j}}.
\ee
The time evolution of our original states is computed by applying the
exponential of the truncated Hamiltonian to the state in question; i.e.,
$
    \ket{\sigma,\bar{x}}(t) = e^{-i {\cal H} t} \ket{\sigma,\bar{x}}.
$
Computing the exponential of the truncated operator is quite simple, except
for one subtlety: we have
defined ${\cal H}$ by its matrix elements between a non-orthonormal set of states.
Hence, to perform the exponentiation, we first
find the eigenvectors and eigenvalues of the metric $N_{ij}$ and use
them to compute the matrix $N^{-1/2}_{i,j}$. \endnote{ If our original set
of states is not linearly independent, then $N_{i,j}$ will have some
zero eigenvalues.  Clearly, we throw their corresponding eigenvectors
away when computing $N^{-1/2}_{i,j}$.  In practice we discard
all vectors whose eigenvalue is smaller than $10^{-5}$.}
Then we construct the transformed ${\cal H}$ by
\be
{\cal H}^{tr}_{i,j} = \sum_{k,l}N^{-1/2}_{i,k}\, {\cal H}_{k,l} \,N^{-1/2}_{l,j} .
\ee
Now we can construct the exponential of this operator by simply finding
its eigenvectors and eigenvalues.
In order to compute the time evolution of one of the original
states we simply write them in terms of the orthonormal basis.

The only step which remains is to explain how we compute the
expectation values of the operator $x$ as functions of time:
we first construct, for each component,
the operator
\be
   {X}_{i,j} = \bra{\sigma,\bar{x_i}} \, x\,\ket{\sigma,\bar{x_j}}
\ee
and use $N^{-1/2}_{i,j}$ to put this into the same basis in which
we exponentiate ${\cal H}$; i.e., construct
\be
    {X}_{i,j} = \sum_{k,l} N^{-1/2}_{i,k}\, X_{k,l} \,N^{-1/2}_{l,j} .
\ee


\end{document}